\documentclass[sigconf,authorversion]{acmart}
\AtBeginDocument{%
  }

\copyrightyear{2026}
\acmYear{2026}
\acmConference[CHI '26]{Proceedings of the 2026 CHI Conference on Human Factors in Computing Systems}{April 13--17, 2026}{Barcelona, Spain}
\acmBooktitle{Proceedings of the 2026 CHI Conference on Human Factors in Computing Systems (CHI '26), April 13--17, 2026, Barcelona, Spain}
\acmPrice{}
\acmDOI{10.1145/3772318.3790286}
\acmISBN{979-8-4007-2278-3/2026/04}

\usepackage{xspace}
\usepackage{listings}
\usepackage{array}
\usepackage{multirow}
\usepackage{soul}

\newcommand{\ie}{{i.e.,}\xspace}
\newcommand{\eg}{{e.g.,}\xspace}
\newcommand{\cf}{{c.f.}\xspace}
\newcommand{\ea}{{et~al.}\xspace}

\newcommand{\etc}{{etc.}\xspace}

\newcommand{\bpstart}[1]{\vspace{1mm} \noindent{\textbf{#1.}}}

\definecolor{codeGreen}{cmyk}{0.85,0.2,1,0.15}
\definecolor{darkblue}{cmyk}{0.95,0,0,0.5}

\lstdefinelanguage{DRACO}{
  keywords={preference, helper, violation, entity, attribute},
  keywordstyle=\color{blue},
  identifierstyle=\color{codeGreen},
  sensitive=false,
  comment=[l]{\%},
  commentstyle=\color{darkgray}\ttfamily,
  stringstyle=\color{darkblue}\ttfamily,
  numberstyle=\color{darkblue}\ttfamily,
  morestring=[b]',
  morestring=[b]"
}

\begin{document}

\title{Automatic Synthesis of Visualization Design Knowledge Bases}

% Authors
\author{Hyeok Kim}
\orcid{0000-0003-4340-4470}
\affiliation{%
  \department{Paul G. Allen School of Computer Science \& Engineering}
  \institution{University of Washington}
  \city{Seattle}
  \state{Washington}
  \country{USA}
}
\email{hyeokk@uw.edu}

\author{Sehi L’Yi}
\orcid{0000-0001-7720-2848}
\affiliation{%
  \department{Department of Biomedical Informatics}
  \institution{Harvard Medical School}
  \city{Boston}
  \state{Massachusetts}
  \country{USA}
}
\email{sehi_lyi@hms.harvard.edu}

\author{Nils Gehlenborg}
\orcid{0000-0003-0327-8297}
\affiliation{%
  \department{Department of Biomedical Informatics}
  \institution{Harvard Medical School}
  \city{Boston}
  \state{Massachusetts}
  \country{USA}
}
\email{nils@hms.harvard.edu}

\author{Jeffrey Heer}
\orcid{0000-0002-6175-1655}
\affiliation{%
  \department{Paul G. Allen School of Computer Science \& Engineering}
  \institution{University of Washington}
  \city{Seattle}
  \state{Washington}
  \country{USA}
}
\email{jheer@uw.edu}

%% shorthand
\renewcommand{\shortauthors}{Kim et al.}

\begin{abstract}
Formal representations of the visualization design space, such as knowledge bases and graphs, consolidate design practices into a shared resource and enable automated reasoning and interpretable design recommendations.
However, prior approaches typically depend on fixed, manually authored rules, making it difficult to build novel representations or extend them for different visualization domains.
Instead, we propose data-driven methods that automatically synthesize visualization design knowledge bases.
Specifically, our methods (1) extract candidate design features from a visualization corpus, (2) select features forward and backward, and (3) render the final knowledge base.
In our benchmark evaluation compared to Draco 2, our synthesized knowledge base offers general and interpretable design features and improves the accuracy of predicting effective designs by 1--15\% in varied training and test sets.
When we apply our approach to genomics visualization, the synthesized knowledge base includes sensible features with accuracy up to 97\%, demonstrating the applicability of our approach to other visualization domains.
\end{abstract}




\begin{teaserfigure}
  \includegraphics[width=\textwidth]{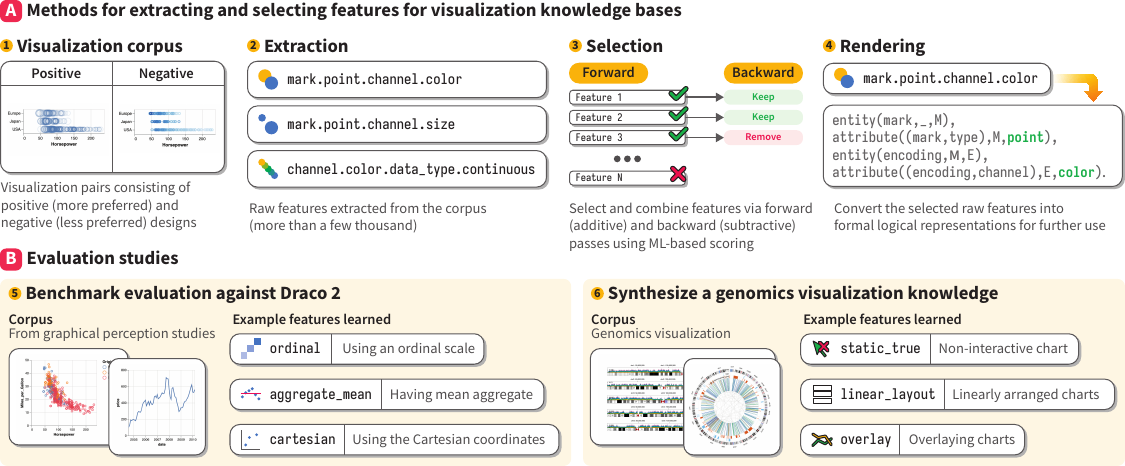}
  \caption{Given a visualization design corpus (1), our methods extract raw features (2), select some of them through forward and backward iterations (3), and then render the selected features into formal representations for further computation, such as visualization design recommendations (4). We evaluate our methods via a benchmark study (5) against an existing visualization design knowledge base (Draco 2~\cite{yang2023:draco2}) and by synthesizing a novel genomics visualization knowledge base from scratch (6).}
  \label{fig:teaser}
  \Description{A. Methods for extracting and selecting features for visualization knowledge bases. First, our methods use input visualization corpus with ordinal pairs each of which one is labeled to be more preferred or positive and the other to be less preferred or negative. Second, our methods extract raw features from the corpus such as mark-point-channel-color. Third, our methods use forward and backward selection methods to select a subset from the raw features. Fourth, our methods render those features into formal representations. B. We apply our methods to a benchmark study against Draco 2, and a scenario for synthesizing a genomics visualization knowledge base. First, using charts from graphical perception studies, our method selected features like using an ordinal scale, using a mean aggregate, and using the Cartesian coordinates. Second, using a genomics visualization corpus, our methods selected features like non-static views, linear arrangement of charts, and overlay of marks.}
\end{teaserfigure}

%% document
\maketitle

\section{Introduction}\label{sec:intro}

Data visualization research has sought formal representation of effective designs to achieve a collective understanding.
Prior work uses representations such as knowledge bases and knowledge graphs to enable useful applications such as explainable automated design recommenders.
For example, Draco~\cite{moritz2018:formalizing,yang2023:draco2} encodes \textit{expressiveness} and \textit{effectiveness} criteria~\cite{mackinlay1986:APT} as logical constraints with weight terms, which allows for ranking designs.
Similarly, KG4Vis~\cite{li2022:kg4vis}, an automated design recommender, employs a knowledge graph where its nodes represent design properties and data characteristics.
Commonly, these approaches represent design knowledge as a feature space (a set of meaningful design properties), though their expression methods differ (\eg~constraints or graph nodes).

A main bottleneck in extending and reusing such knowledge representations is their use of a fixed feature space.
For instance, Draco requires hand-crafted logical constraints, and KG4Vis relies on a fixed set of knowledge nodes from prior work. 
While we can update the weight terms for Draco-like systems or the graph embedding for KG4Vis-like systems using machine learning (ML) techniques, we cannot learn new constraints or knowledge nodes. 
This fixity has consequences: for example, Draco 2's constraints cannot distinguish designs in four out of 30 graphical perception studies collected by Zeng~\ea~\cite{zeng2024:tooManyCooks}.
A fixed feature space further makes it difficult to build intelligent authoring systems for other visualization domains (genomics, networks, \etc), as we first need to build a feature space from scratch and then manually assess the soundness and relevance of individual features---a tedious and error-prone process. 
We need a more generic method of building knowledge representations for effective visualization design. 

Our work proposes an automated approach to synthesizing a knowledge base for effective visualization design using a corpus of visualizations. 
Our method first extracts low-level design facts (\eg~using a point mark) from a corpus and combines them to form more complex design features (\eg~using \textit{x} and \textit{color} channels for point mark).
As this process can produce tens of thousands of features, we select a much smaller subset through forward and backward feature selections, where features are evaluated in terms of their performance in predicting effective visualization designs.
During forward selection, we add each feature in the order of their estimated ability to determine effective designs and generalize to other designs using multiple proxy metrics. 
As forward selection can at times add relatively less useful features before adding some later useful features, the backward selection step tests the effect of omitting features in a random order.
Lastly, we convert the selected features into a logical format, over which constraint solvers can perform logical reasoning. 

We evaluate our method in two experiments: (1) a benchmark comparison with Draco 2~\cite{yang2023:draco2} and (2) automatic synthesis of a knowledge base for genomics data visualizations.
In the benchmark study, our method provides comparative accuracy in predicting the effectiveness of visualization designs from prior work~\cite{mackinlay1986:APT,saket2018:encodings,kim2018:encodings} that informed the original Draco feature space (about 93--94\%) using fewer features.
When considering additional designs, our method improves performance by 4--5\%, achieving 80--89\% prediction accuracy.
We further find that some design features in Draco 2 can be combined into a single feature with less distinctions or concatenated with more precision.
For example, Draco 2 distinguishes designs using a continuous size channel by different task types, while our method selects a feature that does not have the same distinction. 
Instead, our method selects a feature using a size channel for point marks, which exist as separate features in Draco 2. 
Our knowledge base for genomics visualizations (expressed using Gosling~\cite{lyi2022:gosling}) achieves an accuracy of up to 97\% in the same prediction task using a visualization corpus we collected.
The synthesized design features capture key design criteria for genomics visualizations with respect to interactivity, layout, and data transformation that reflect characteristics of genomic data, such as enormous data size and need for parallel comparison.
We solicited feedback on the selected features from two genomic data experts, and they in general found the selected features to be helpful in reasoning about genomic visualization designs.

To summarize, our contributions include: 
\begin{itemize}
    \item An automated method for synthesizing knowledge bases for effective visualization design from a corpus of chart pairs,
    \item A benchmark study against an existing knowledge base (Draco~2) with prediction accuracy up to 94\%, and
    \item Automatic synthesis of a novel genomics visualization knowledge base with prediction accuracy up to 97\%.
\end{itemize}
\section{Related Work}\label{sec:rw}

Our work is grounded in prior research on knowledge-based visualization design recommendation as well as feature selection techniques for machine learning. 

\subsection{Knowledge-based Visualization Design Recommendation and Reasoning}
Much prior work on automated visualization recommendation leverages knowledge representations, pursuing informative and explainable reasoning about effective designs.
For instance, Mackinlay's APT~\cite{mackinlay1986:APT} includes criteria for desired and grammatically correct designs (\textit{effectiveness} and \textit{expressiveness} criteria).
Positional encodings (\textit{x} and \textit{y}) are more likely to produce \textit{effective} designs than non-positional ones (\textit{shape} and \textit{size}); yet encoding a categorical field using a continuous scale results in an \textit{inexpressive} design that can confuse readers. 
Inspired by this early work, there have been two major approaches in knowledge-based visualization recommendation: using either models or knowledge bases.
These approaches share abstract-level commonalities, such as describing design properties as design ``features'' and using recommendation engines that assign weight terms to those design features.
We use features to generally refer to meaningful design properties that support reasoning about different designs.
Features can range from simple attributes (\eg~having an \textit{x} channel) to complex combinations (\eg~using non-positional channels without any positional channels).
Because features themselves cannot rank different designs, they have numerical \textit{weight terms} that indicate their desirability or preference.
In this way, recommendation engines can consider varying importance and sign (negative/positive) of design features.

Model-based approaches train machine learning (ML) models to recommend a few visualization designs given data characteristics.
For example, VizML~\cite{hu19:vizml} relies on a set of features representing a prior understanding of visualization design, such as mark type choices, having a shared axis, and the entropy of a variable.
To address the difficulty in associating relationships between those atomic features, KG4Vis~\cite{li2022:kg4vis} uses a numerical embedding for a knowledge graph that connects data features (\eg~entropy) and design features (\eg~using a bar mark).
A machine learning (ML) training associates a weight term to each link between a pair of data and design features.
AdaVis~\cite{zhang2024:adavis} is another model-based approach that leverages a knowledge graph with a feature set from prior work. 
Yet, these approaches tend to result in model parameters or numerical embedding matrices, which limit the interpretability and reuse of learned knowledge.

In contrast, approaches leveraging a knowledge base have recommendation engines composed of explicitly written formal inference rules as features.
Extending APT~\cite{mackinlay1986:APT}, systems like BOZ~\cite{casner1991:boz}, Visage~\cite{schroeder1992:visage}, Vista~\cite{senay1994:vista}, SAGE~\cite{roth1988:sage1,mittal1998:sage}, CompassQL~\cite{wongsuphasawat2016:compassQL}, and Draco~\cite{moritz2018:formalizing,yang2023:draco2} encode a set of rules to produce desired and plausible designs.
Specifically, Vista and Draco offer expression methods that consistently capture visualization designs and effectiveness conditions.
This consistency supports tasks beyond recommendation, including completing partial design specifications and automated reasoning about existing designs.
Useful applications, such as visualization authoring tools (\eg~Voyager~\cite{wongsuphasawat2016:voyager,wongsuphasawat2017:voyager} and Dupo~\cite{kim2023:dupo}), have employed knowledge bases for iterative design recommendations.
In genomics visualization, GenoREC~\cite{pandey2023:genorec} uses a set of rules based on prior design space research~\cite{nusrat2019:genomic}.

Our work extends this knowledge base-based approach, adopting expressions from Draco 2~\cite{yang2023:draco2} given its flexibility and explicitness.
Draco 2's minimal and recursive expressions allow for representing design properties beyond predefined ones.
Plus, the explicit encoding supports auditing its reasoning via inspection of both features and their weights~\cite{schmidt2024:dracova, wang2025:dracogpt}.

\subsection{Updating Visualization Knowledge Bases}
To update a visualization knowledge base, prior work focused on updating weight terms using ML-based techniques.
Model-based systems use trained models as recommendation engines. 
Knowledge bases can also train a relatively simple model (typically, logistic regression or support vector machines) and apply the learned coefficients as weight terms for the corresponding features. 
A major challenge in updating weight terms is to find an exhaustive corpus that can sufficiently cover the feature set. 
To enhance this process, Kim and Heer~\cite{kim2025:augmentation} recently proposed augmentation methods to generate novel training data to improve weight term updates.

Yet, prior work on knowledge-based visualization recommendation relies on a fixed set of design features.
For instance, Draco~\cite{yang2023:draco2,moritz2018:formalizing} uses logical constraints as features manually drawn from prior work.
Similarly, KG4Vis~\cite{li2022:kg4vis} and AdaVis~\cite{zhang2024:adavis} use manually collected features in addition to the feature set from VizML~\cite{hu19:vizml}.
This fixed nature makes it difficult to expand a knowledge base to cover cases beyond its original scope.
For example, our internal audit with Draco found nine pairs of stimuli from four graphical perception studies (out of 1,384 pairs from 30 studies collected by Zeng~\ea~\cite{zeng2024:tooManyCooks,zeng2023:dataset}) that Draco cannot distinguish, yet corresponding experiment results indicate that they are highly relevant. 
Although it is a relatively small number, those designs were curated to test specific and important design features, such as overplotting. 

In addition, it is tedious and error-prone to manually craft expressions for knowledge bases without training in logic programming. 
This technical difficulty can further hamper the wide adoption of knowledge bases for other specialized design areas, such as natural science, network analysis, or decision under uncertainty, where domain practitioners are not necessarily trained in visualization design as well as logic programming. 
Considering the above needs for feature set expansion and applications to other visualization domains, our work contributes a method for automatically synthesizing a knowledge base from a corpus of charts. 
In addition to a benchmark evaluation, we demonstrate the usefulness and generalizability of our method by applying them to genomics visualization. 

\subsection{Feature Selection in Machine Learning}
The number of features may explode when extracted from data.
However, we cannot apply all of them because it increases the computation complexity (more features than data points) and leads to overfitting. 
Furthermore, some raw features may target aspects of individual cases that are too specific (\eg~encoding a temporal field while the chart includes two or more non-positional encoding channels), and hence do not generalize or support broader interpretability.
To prevent such unwanted situations, it is wise to collect a subset of features that most contribute to performance and interpretability.
Given that both of the above approaches use ML-based methods to learn weight terms, we take inspiration from ML feature selection methods. 
As it is a fast-growing area, we only review those relevant to our work (supervised classification learning).
Further references can be found in review articles~\cite{theng2024:feature,khalid2014:feature}.

Supervised learning of classifiers has used \textit{filters} and \textit{wrappers} for feature selection~\cite{theng2024:feature}. 
\textit{Filters} are certain metrics that estimate the success of candidate features in terms of information gain and correlation to the dependent variable, for example. 
Models can then choose a certain number of top-ranked features using filters. 
While this reduces the computation costs, it may fail to consider features that are important but less represented by the dataset~\cite{maryam2018:filter}. 
A \textit{wrapper} is an iterative process that assesses the contribution of each feature to the prediction performance~\cite{sanz2018:wrapper}.
A feature is selected if it increases the performance.
Forward and backward selection methods fall into this category. 
Wrappers can lead to higher performance because of their exhaustiveness with the caveat of potential overfitting.
As a hybrid method, embedded techniques take the advantages of both approaches (\eg~\cite{pes2020:embedded}).
In general, embedded techniques rank features using some filters and then feed them to a wrapper in that ranked order.
We apply this embedded technique for our forward and backward feature selection steps.

\begin{figure}
    \centering
    \includegraphics[width=\linewidth]{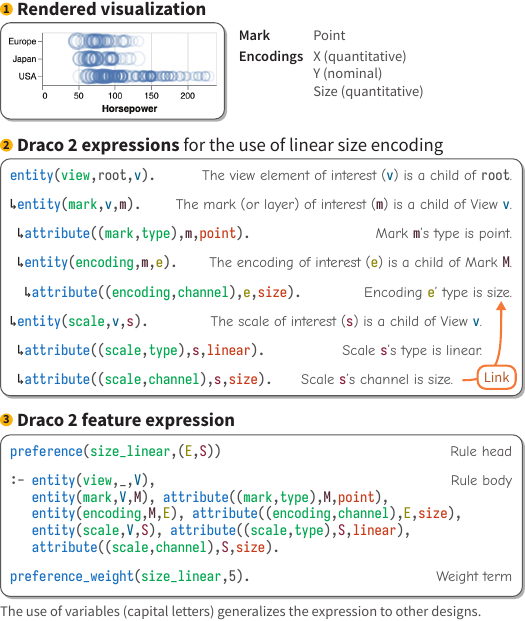}
    \caption{How Draco~2~\cite{yang2023:draco2} expresses a visualization design and captures a feature.}
    \label{fig:draco2intro}
    \Description{1. An example visualization that has point marks with the quantitative x, nominal y, and quantitative size channels. 2. Draco 2 expressions for the linear size encoding for the above chart. 3. Draco 2 feature expression for the same characteristic. But this time, the expression uses variables to generalize the feature to other designs.}
\end{figure}

\subsection{Technical Background for Draco 2}\label{sec:draco2}

Draco 2~\cite{yang2023:draco2} (hereafter, simply Draco) employs Answer Set Programming (ASP)~\cite{lifschitz2008:asp} to encode design features as logical constraints using predicate logic.
To show how Draco expresses a visualization design and captures its design features, suppose that we are capturing the linear size encoding in a dot plot (\autoref{fig:draco2intro}-1).
Draco first defines visualization elements and their parents (membership) using \lstinline{entity} statements.
As shown in \autoref{fig:draco2intro}-2, for example, \lstinline{entity(encoding,m,e).} means the entity \lstinline{e} is an \lstinline{encoding} and its parent is \lstinline{m} (mark/layer).
Using this membership information expressed as entity identifiers like \lstinline{e} and \lstinline{m}, logic solvers can avoid considering unrelated elements by mistake, which is a common practice in logic programming.
Then, an \lstinline{attribute} statement adds details about an element. For instance, \lstinline{attribute((encoding,channel),e,size).} indicates that the \lstinline{encoding} \lstinline{e}'s \lstinline{channel} is \lstinline{size}.
This entity-attribute structure allows for expressing any tree-structured design specs.

However, the expressions in \autoref{fig:draco2intro}-2 describes the design property of a given design specification only.
If the same design uses different entity identifiers (\eg~\lstinline{v1} instead of \lstinline{v}), the same expressions are no longer applicable.
To detect the same design feature from a larger set of designs, Draco employs generalized expressions using variables (in capital letters) as shown in \autoref{fig:draco2intro}-3. 
The \lstinline{root} element is no longer relevant as long as the mark and scale belong to the same view. In this case, we drop the identifier with a dummy marker \lstinline{_}.
We can then wrap the generalized version as a design feature using a logical rule.
Here, the rule head is \lstinline{preference(size_linear,(E,S)).} where \lstinline{size_linear} is the feature's name and \lstinline{(E,S)} are identifiers for tracing purposes. The rule body contains the generalized expressions.
Finally, to represent the desirability or importance of a feature, Draco assigns a weight term to each feature.
In this case, \lstinline{preference_weight(size_linear,5)} means a weight of $5$ for each occurrence of the \lstinline{size_linear} feature.
Given a design specification, a logic solver (\eg~Clingo~\cite{gebser2014:clingo} for ASP) detects features, assigns the corresponding weights, and takes the weight sum, which can then be used to rank a set of designs.

To update Draco weights, we can use a visualization corpus consisting of ordered design pairs.
In each pair, the preferred design is labeled \textit{positive}, and the other is \textit{negative}.
For example, \autoref{fig:teaser}-A1 illustrates a case where a quantitative size encoding works better than a quantitative color encoding, according to prior work~\cite{mackinlay1986:APT,kim2018:encodings}.
In the Draco feature space, both designs share most features but two:  \lstinline{size_linear} in the positive chart and \lstinline{color_linear} in the negative one. 
If we train a classifier (\eg~SVM or logistic regression) on the design corpus (expressed as design features), the model will learn to penalize \lstinline{color_linear} and encourage \lstinline{size_linear}, producing commensurate coefficients for those features.
We then use those coefficients as updated weights for those features.
To evaluate the updated weights, we measure their accuracy in predicting pairwise preferences.
We use this process as a wrapper in the selection steps, and provide further relevant details as we describe our method.
\section{Guidelines}
We applied the following guidelines while developing our method for automatic knowledge base synthesis.

\bpstart{G1. Features should be both Sensible and Useful}
First, the knowledge representations should be clear and interpretable to human users and researchers. 
If synthesized features are overly specific to a degree where a feature tightly corresponds to an individual chart, it is less helpful to inspect what contributes to better visualization design (as noted by prior work~\cite{hu19:vizml,li2022:kg4vis,yang2023:draco2}).
Therefore, synthesis methods need to export meaningful, interpretable, and sensible features to support human reasoning about visualization designs.
Towards this aim, we use selection filters (metrics) that estimate the relevance between the elements in a feature \textit{structurally} (their distance in an abstract syntax tree) and \textit{distributionally} (their co-occurrence in a corpus).
Later, we also qualitatively review selected features and compare with prior approaches.

On the other hand, synthesized features need to effectively support valuable applications of design knowledge bases, such as automated design reasoning and recommendation. 
For instance, consider a feature that a mark type (of any type) is defined. 
This feature is highly interpretable, yet it is likely to have no power for distinguishing effective designs, as it would apply to all charts.
Synthesis methods must consider the usefulness of features for identifying effective visualization designs, which we assess using accuracy in predicting pairwise preference.

\bpstart{G2. Feature Construction should be Configurable and Adoptable}
There can be features that encompass multiple visualization elements in a relatively complex way.
For instance, Draco has simpler features such as using a log scale for the \textit{x} channel, as well as more complex features such as avoiding overlap between bar marks using stacking, aggregation, or binning. 
Instead of entirely relying on inductive extraction, it makes sense to offer a way to configure such complex features. 
Thus, our synthesis process includes systematic methods to encode \textit{a priori} knowledge.

At the same time, knowledge base synthesis methods should be applicable to novel or specialized visualization areas, such as genomics, network, or uncertainty visualization.
If synthesis methods are hard-coded, it might be difficult to modify them to reflect design characteristics specific to a target domain.
For example, in genomics visualization, there is a higher degree of freedom in configuring multiple views while the data structure is somewhat fixed. 
In contrast, communicative visualizations require higher flexibility in encoding choices and data structures, but they seldom use a lot of views in a single chart. 
To ensure broad adoption of automated reasoning about effective visualization through knowledge bases, synthesis methods must be configurable to reflect domain-specific design constraints.
We demonstrate the configurability and adoptability of our method by applying them to genomics visualization.
\section{Knowledge Base Synthesis Method}\label{sec:method}

Our knowledge base synthesis method produces a set of formally represented design features for a knowledge base as the final output.
As illustrated in \autoref{fig:teaser}-A, this process comprises the following three parts.
First, we \textbf{extract} candidate features from input data consisting of visualization pairs (\autoref{fig:teaser}-A2), which results in thousands of candidates.
Next, we \textbf{select} feasible features from the enumerated features, producing a much smaller set of features (\autoref{fig:teaser}-A3).
Lastly, we \textbf{render} the selected features as formal representations for inclusion in a knowledge base (\autoref{fig:teaser}-A4).
In this section, we describe our method at a high level. 
Source code and additional technical details are provided in the Supplementary Material for replicability.

\subsection{Assumptions about Input Visualization Data}

Our method has the following requirements for processing purposes, while not overly restricting the applicable design space.
First, we need a corpus of ranked visualization pairs where one design is labeled as preferred to the other; we refer to these as \emph{positive} and \emph{negative} charts (\eg~\autoref{fig:teaser}A1). 
For example, experimental stimuli can be ordered based on their effectiveness rankings, as collected by Zeng~\ea~\cite{zeng2023:dataset,zeng2024:tooManyCooks}.
When confirmatory empirical data is not available, LLMs may provide feasible initial labels, as recent work demonstrates~\cite{wang2025:dracogpt,kim2025:augmentation}.
Using ordered pairs helps us to learn about design features that are contributing to ``less preferred'' designs as well as ``more preferred'' ones. 
Next, visualization pairs need to be expressed in a specification format, such as Vega~\cite{satyanarayan:vega-lite2017,satyanarayan:vega2016,vanderplas:altair}, ggplot2~\cite{wickham:ggplot22010}, or Matplotlib~\cite{hunter:matplotlib}.
As many modern visualizations are generated using declarative grammars and given advances in object retrieval methods in visualization~\cite{wu21:lq2}, this should not significantly confine the scope of applicable visualization data.

\subsubsection{Step 1. Prepare the data}
Given a visualization design corpus formatted in the above way, we convert it into formal representations for consistency in the later steps.
Specifically, we adopt the representation method of Draco~\cite{yang2023:draco2} (\autoref{sec:draco2}).
We choose this representation method because it allows for serializing any mark-based and encoding-based visualization spec with a hierarchical structure (\ie~expressible using JSON).
In addition, we can keep track of which element an attribute belongs to, which is important information in the feature combination step.
It is also possible to define custom entity and attribute types, which provides further flexibility to apply our method to different visualization domains (\textbf{G2}: Configurability \& Adoptability).
While Draco offers a compiler that converts a Vega-Lite spec into Draco expressions, other declarative grammars, such as Gosling~\cite{lyi2022:gosling} or ggplot2~\cite{wickham:ggplot22010}, can be supported by implementing a separate compiler.
For example, we implemented a Gosling-to-Draco compiler for the genomics visualization evaluation in \autoref{sec:eval:genomics}. 

\begin{figure}
    \centering
    \includegraphics[width=\linewidth]{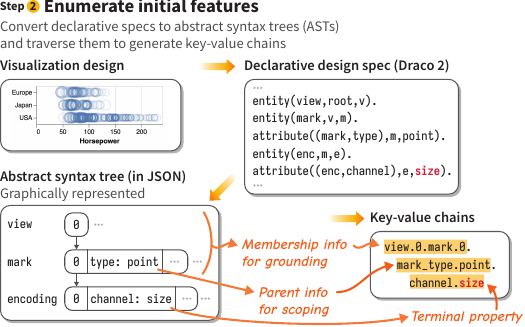}
    \caption{Enumerating initial features (key-value chains) from a declarative design spec. The identifiers in a Draco specification are replaced with numerical indices when converted to an abstract syntax tree.}
    \label{fig:step2}
    \Description{Steps for enumerating initial features. First, we convert declarative specs to abstract syntax trees in JSON. Then, we traverse those abstract syntax trees to generate key-value chains that consist of membership information for grounding, parent information for scoping, and terminal property.}
\end{figure}

\subsection{Extraction}

Given design pairs written in the Draco format, we extract candidate features. 
The main goal of this stage is to enumerate as many sensible features as possible, while also pruning the search space to make the following selection steps both informative and efficient.
As we extract and combine features, we compute their frequency vectors along the way, as illustrated in \autoref{fig:frequency}.

\subsubsection{Step 2. Enumerate initial features (key-value chains)}

We traverse each (not pair) visualization design to populate the initial, unary features.
As illustrated in \autoref{fig:step2}, Draco specs form an abstract syntax tree (AST), so we run depth-first traversal and produce key-value chains.
For example, traversing to a size encoding channel attribute, \lstinline{attribute((encoding,channel),m,size).}, we get a key-value chain of \lstinline{view.0.mark.0.mark_type.point.channel.size}.
After traversing one step further down, we obtain another key-value chain of \lstinline{view.0.mark.0.mark_type.point.channel.size.data_type.continuous}.
Here, \lstinline{view.0.mark.0} is membership information (only for grouping entities), \lstinline{mark_type.point.channel.size} is parent information, and \lstinline{data_type.continuous} is the terminal key and value.
Note that a mark type and an encoding channel appear only once for a layer, regardless of the visualization grammar, so they are always included as parent information.
This treatment can reduce the search space by removing combinations like \lstinline{channel + data_type} while preserving parent information for channel-specific combinations like \lstinline{scale_type + data_type}. 

When terminal values are numerical (\eg~entropy, unique values), having each value as a separate key-value chain only increases the search space without information gain due to the high specificity.
Thus, we collect such numerical values and set boundaries for them, so that the final knowledge base can have features like `having a discrete color channel for more than 15 unique values.'
For example, the number of unique values (ranging from 0 to 35) can have boundaries, such as $[0,15)$, $[15,25)$, and $[25,35)$.
We determine these boundaries using 1D k-means clustering. 
We iteratively cluster the numerical values for each terminal key (two to ten clusters) and choose the best clustering based on the silhouette score~\cite{shahapure:silhouette}. 
We adjust those boundary values to human-friendly values (\eg~281 to 250) to the extent that they still preserve the clustering.

\begin{figure*}
    \centering
    \includegraphics[width=\textwidth]{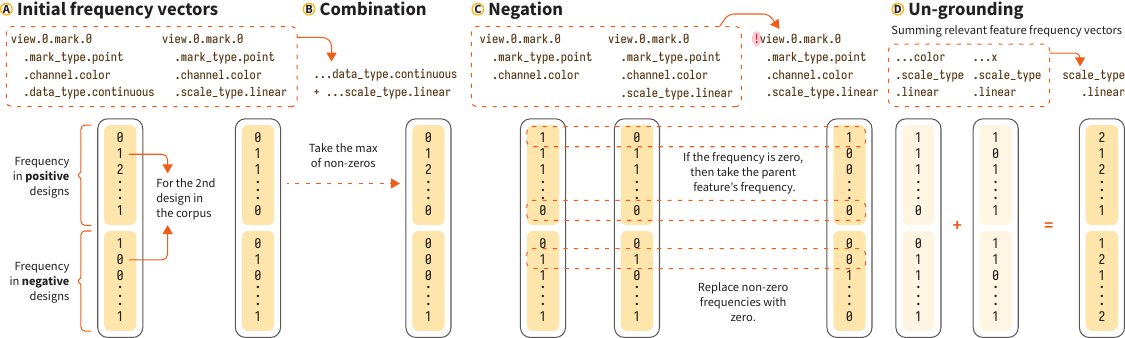}
    \caption{Computing frequency vectors through the extraction steps.}
    \label{fig:frequency}
    \Description{We compute feature frequency vectors to combine, negate, and un-ground features. A frequency vector has two sections, one for positive designs and the other for negative designs. For combination, we take the maximum value of each element in the feature vectors to be combined, only when all of them are non-zero. For negation, we covert the non-zero values for the feature vector to be negated to zeros, only when the corresponding values in its parent frequency vector (if exists) are non-zero. Please refer to the paper for details.}
\end{figure*}

The output of this step is the frequency vectors for those initial features (\autoref{fig:frequency}A).
For example, the $i$-th element in the frequency vector for \lstinline{view.0.mark.0.mark_type.point.channel.size} is its frequency in the positive design of the $i$-th pair in the corpus, and the $(2 * i)$-th element is its frequency in the negative design. 

At this stage, a developer can provide some \emph{a priori} knowledge that is widely known. 
For example, different combinations of position encodings (\eg~continuous by continuous, discrete by discrete) are key information for determining a mark type, which may not appear quite clear when looking at individual key-value chains. 
Thus, it is sensible to allow for providing known important design properties, further reducing the search space in later steps.  
For example, it is widely known that having too many encoding channels adds visual complexity.
Our implementation includes helper functions to easily encode a priori knowledge in a JSON format, which further produces logical inference rules (\textbf{G2: Configurability}). 
To encode the previous example, we provide a helper function to define the frequency of a certain property (\ie~\lstinline{encoding} entities in this case).

\begin{figure}
    \centering
    \includegraphics[width=\linewidth]{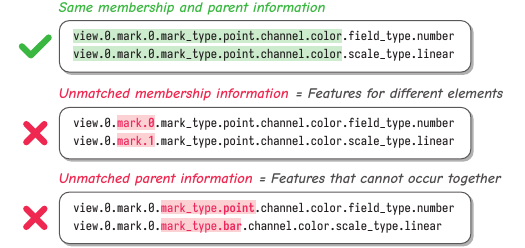}
    \caption{Conditions for combining initial features.}
    \label{fig:step3}
    \Description{Two key-value chains share the same membership and parent information, which is a chain of view-0, mark-0, mark type of point, and channel of color. They are eligible for combination. However, features with different membership information are for different visualization elements, and features with different parent information cannot occur together, so they are excluded for combination.}
\end{figure}

\subsubsection{Step 3. Combine and negate features}
Given that the previous step did a depth-first traversal, we also need to perform a breadth-wise search to capture combinations of design features.
An important consideration here is to scope feasible combinations. 
As shown in \autoref{fig:step3}, two key-value chains for the same encoding channel and the same mark type can be combined as they share membership and parent information.
However, it is less desirable to combine features for different layers (membership information) or different mark types (parent information) as they are irrelevant. 
On the other hand, a combination of higher-level and lower-level features can be useful.
For example, having an \textit{x} channel (encoding-level) when there are only two encoding channels (view-level) might reflect a general preference for position encoding channels over non-positional ones. 
Our implementation offers utility functions for configuring different scoping rules.
Lastly, we check whether the intersection of the frequency vectors of candidate feature combination is non-zero, in order to exclude uninformative combinations.

We then compute the feature frequency vectors of the plausible feature combinations.
As shown in \autoref{fig:frequency}, consider two combinable vectors $a$ and $b$. 
When the $i$-th elements, $a_i$ and $b_i$, are non-zero, their maximum value is used in the new, combined frequency vector.

In doing so, we also negate a subset of each combination to promote or penalize cases of ``not doing X given Y.''
For example, consider a combination of \lstinline{...channel.x.data_type.continuous} (having a continuous \textit{x} encoding) and \lstinline{...channel.x.scale_type.linear} (having a linear scale \textit{x} encoding).
We can negate either one of them (but not both), producing two combinations: \lstinline{!...data_type.continuous} + \lstinline{...scale_type.linear} and \lstinline{...data_type.continuous} + \lstinline{!...scale_type.linear}, where \lstinline{!} indicates negation.
To compute the frequency vector for a negated feature (\autoref{fig:frequency}-C), we first replace non-zero elements in the original frequency vector with zeros.
For each zero element in the original vector, we take the corresponding element in the parent vector.
For example, if the $i$-th element of the  \lstinline{...channel.x.data_type.continuous} vector is zero, then we take the $i$-th element of the  \lstinline{...channel.x} vector (parent information vector) for the corresponding element of the \lstinline{!...channel.x.data_type.continuous} vector.
In this way, we avoid logically invalid cases, for example, where a design has no \textit{x} channel but has a feature for not having a continuous \textit{x} channel. 

\begin{figure}
    \centering
    \includegraphics[width=\linewidth]{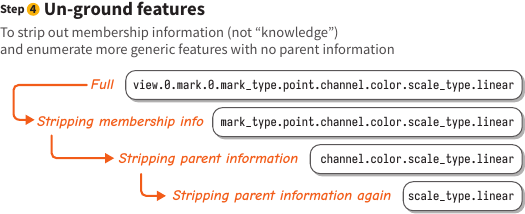}
    \caption{Un-grounding features by stripping out membership and parent information.}
    \label{fig:step4}
    \Description{We un-ground features by stripping out membership and parent information recursively to produce more generic features. For example, suppose a chain of view-0, mark type of point, channel of color, and scale type of linear. Then, we first remove view-0 and mark-0, as they are membership information and hence do not contribute to the feature itself. We can further remove mark type of point, and then channel of color, resulting in the final, generalized feature for scale type of linear.}
\end{figure}

\subsubsection{Step 4. Un-ground features}

Up to this stage, key-value chain features have membership information that is important for combining features, but is not specific to design characteristics.
By additionally removing parent information, we can test if a feature is generalizable to a larger set of designs. 
As shown in \autoref{fig:step4}, we remove membership information first, then iteratively remove parent information.
Consider an un-grounded feature \lstinline{scale_type.linear} sourced from \lstinline{channel.color.scale_type.linear} and \lstinline{channel.x.scale_type.linear}.
To compute the new frequency vector (\autoref{fig:frequency}-D), we take the sum of the original feature frequency vectors.
For combined features, we strip the common membership and parent information only.
For instance, when combining \lstinline{...mark_type.point.channel.x.bin.true} and  \lstinline{...mark_type.point.channel.y.aggregate.count}, our method can strip up to \lstinline{mark_type.point} while keeping the channel information.

\subsection{Selection}

Once we collect the frequency vectors of candidate features, we run selection procedures (\autoref{fig:teaser}-A3).
To overview, the forward selection step evaluates one feature at a time and adds it if it improves the pairwise design preference prediction score.
Next, the backward selection step omits a certain number of randomly chosen features at a time.
If the omission improves the performance score, then those features are permanently removed from the selected features.

\begin{figure}
    \centering
    \includegraphics[width=\linewidth]{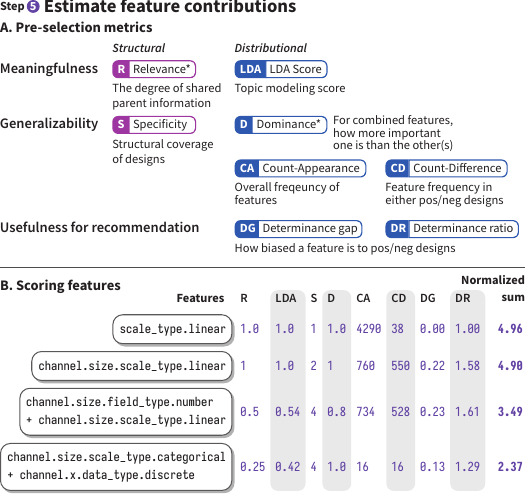}
    \caption{Pre-selection metrics for estimating feature contributions. *Metrics for combined features. For unary features, the value is 1.}
    \Description{A. Pre-selection metrics for meaningfulness, generalizability, and usefulness for recommendation. Meaningfulness has a structural metric of relevance, and a distributional metric of LDA score. Generalizability has a structural metric of specificity, and three distributional metrics of dominance, appearance count, and difference count. Lastly, usefulness of recommendation has two distributional metrics of determinance gap and determinance ratio. B. We score features by taking the normalized sum of the pre-selection metrics.}
    \label{fig:step5}
\end{figure}

\subsubsection{Step 5. Estimate feature contributions}

To select potentially useful features early in the forward selection, we estimate the expected contributions of the collected features using \emph{pre-selection metrics}. 
In particular, we use metrics for meaningfulness, generalizability, and usefulness for prediction quality, as outlined in \autoref{fig:step5}-A.
We have structural (extracted from key-value chains) and distributional (extracted from the corpus) metrics except for the usefulness criterion as it estimates prediction quality based on a corpus.
Here we describe the individual metrics at a high level; refer to the Supplementary Material for a detailed formulation.

First, we need to assess if combined features have a meaningful and sensible relationship (\ie~not a random combination, \textbf{G1}).
We use \textit{relevance} and \textit{LDA} (topic modeling) scores.
For structural meaningfulness, the relevance score computes the degree of shared parent information among the components of each combined feature.
Next, to see the correlation-based meaningfulness, we take the maximum average LDA coefficients of a combined feature.
After modeling topics from un-grounded, unary features (their frequency vectors) using LDA, we obtain their coefficients per modeled topic that represent how cohesive they are in that topic.
Then, we take the average coefficient of the components of each combined feature for the topic that maximizes the average.
We choose LDA because clustering is exclusive and correlation can only be applied to a binary combination.
This distributional LDA score can cover cases where the structural relevance score might ignore.
For example, the relevance score is lower for a combined feature for linear color and linear \textit{x} than one for log and continuous color, yet the former might also be highly important (\cf~\cite{kim2018:encodings}).
For unary features, we assign relevance and LDA scores of 1 to prioritize simplicity.

Next, we wish to evaluate more generalizable features first during forward selection, using \textit{specificity}, \textit{count-appearance}, \textit{count-difference}, and \textit{dominance} scores. 
The specificity score indicates how many cases a feature can cover structurally.
For example, \lstinline{scale_type.linear} covers more designs than \lstinline{channel.size.scale_type.linear}.
We compute the specificity score by taking the sum of the lengths of key-value chains in each feature.
We assess the actual coverage of a feature using the count-appearance score.
At the same time, we also count the number of times that each feature appears differently in positive and negative designs (count-difference). 
For instance, if 75 pairs have \lstinline{scale_type.log} either in their positive or negative designs (but not in both), then the difference count for that feature is 75. 
If a component in a combined feature is more important than the others, it may indicate that the combined feature is overly specified, a potential cause for overfitting.
To address these features, we use the dominance score, defined as the ratio between the smallest and largest importance values of the components of each combined feature.  
For the importance of ungrounded unary features, we train a decision tree with gradient boosting and take the feature importance values.

Lastly, we apply a couple of metrics to estimate the usefulness of a feature for making predictions about preferred designs.
To do so, we first compute the positive and negative relative frequencies of features. 
The positive relative frequency of a feature is defined as its frequency in positive charts divided by the number of pairs in the corpus, and the negative relative frequency is computed likewise.
Then, we compute the (absolute) gap and ratio between positive and negative relative frequencies. 
If the ratio value is lower than 1, we take its inverse to retain a consistent scale. 
A feature with lower determinance has the gap closer to zero and the ratio closer to one. 

We take the normalized sum of the above metrics, and use it to sort the features to add during forward selection.
It is more generalizable to have individual unary features than their combinations, as long as they perform well in reasoning about preferred designs. 
Thus, when sorting features, we use feature dependency information so that the unary components of a combined feature come before their combination.

\subsubsection{Step 6. Initial selection}

During the early phase of forward selection, we may not be able to see top-rated features making an improvement in the preference prediction.
A potential cause can be that the feature set is too small to make any good prediction performance.
To prevent our method from ignoring features that could be useful after adding more, we choose the top N features ranked using the pre-selection metrics to form a baseline selection.
Our backward selection step can later re-assess these initial features.

\begin{figure}
    \centering
    \includegraphics[width=\linewidth]{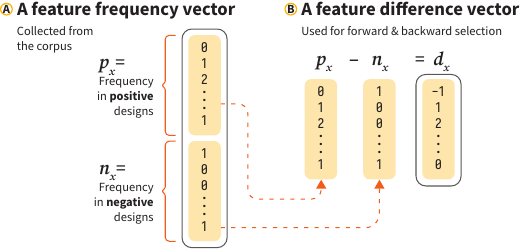}
    \caption{Computing a feature difference vector.}
    \label{fig:feature-diff}
    \Description{We compute a feature difference vector from a frequency vector. We split a feature frequency into two parts: positive and negative. Then we take the difference of those two parts.}
\end{figure}

\begin{figure}
    \centering
    \includegraphics[width=\linewidth]{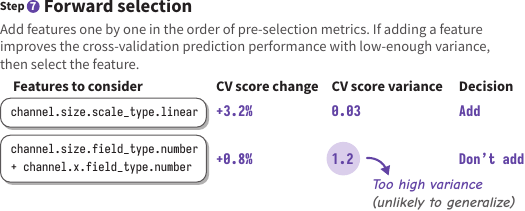}
    \caption{Criteria for adding or ignoring a feature during a forward selection step. The same criteria also apply to backward selection.}
    \label{fig:step7}
    \Description{During a forward selection step, if a candidate feature increases CV score and has CV score variance lower than a predetermined threshold, then we add that feature into the selection. If one of these criteria is not satisfied, then the feature is ignored.}
\end{figure}

%tc:ignore
\begin{table*}
    \caption{Parameter definitions for the selection steps and parameter values used for the evaluation studies. We selected parameter values after several test rounds to see if selection processes terminate too early (\eg~adding 5 features) or too late (\eg~adding several thousand features).}
    \label{tab:parameters}
    \Description{Parameters used for our selection methods. Please refer to Section 4.3 for details.}
    \centering
    \small
    \setlength\extrarowheight{2pt}
    \resizebox{\textwidth}{!}{
    \begin{tabular}{p{2cm}p{2cm}p{7cm}p{2cm}p{2cm}} 
        \textbf{Step} & \textbf{Name} & \textbf{Description} & \textbf{Benchmark} & \textbf{Genomics KB} \\ \hline
        5 (estimate) & Num. topics & The number of topics for LDA 
        (topic modeling) score & 80 & 80 \\ \hline
        6 (initial) & Size & The number of features to select initially (before forward) & 30 & 20 \\ \hline
        7-8 (selection) & \raggedright{Improvement threshold} & Add a feature if it improves the average cross validation accuracy score by more than this value & 0.1\% & 0.1\% \\ \hline
        7-8 (selection) & \raggedright{Variance threshold} & Do not add a feature if the cross validation score standard error exceeds this value & 0.05 & 0.025 \\ \hline
        7 (forward) & \raggedright{Pre-selection metric bound} & Stop forward selection once the sum of pre-selection metrics is below this value & 2 & 2 \\ \hline
        7 (forward) & \raggedright{Convergence threshold} & Count a forward iteration as ``converging'' when the (absolute) performance change is below this value & 0.1\% & 0.1\% \\ \hline
        7 (forward) & \raggedright{Convergence count} & Stop forward selection after getting converging iterations more than this number of consecutive times & 200 & 200 \\ \hline
        7 (forward) & \raggedright{Not selected count} & Stop forward selection after not selecting a feature more than this number of consecutive times & 3,000 & 3,000 \\ \hline
        7 (forward) & \raggedright{Feature set size} & Stop forward selection upon reaching this size of feature set & 500 & 500 \\ \hline
        8 (backward) & \raggedright{Iteration counts} & Iterate backward selection this number of rounds & 10,000 & 10,000 \\ \hline
    \end{tabular}
    }
\end{table*}
%tc:endignore

\subsubsection{Step 7. Forward selection}

This step iteratively adds a feature in the order of the sum of pre-selection metrics (Step 5). 
When adding a feature, we test if the feature can improve prediction for pairwise preferences with cross validation.
Suppose that we add a feature $x$ to the currently selected feature set, $C$. Then, $x$ and $C$ form a new feature set $C' = \{C_0, \cdots, C_n,x\}$, where $C_i$ is $i$-th selected feature.
Then (inspired by Draco-Learn~\cite{moritz2018:formalizing}), we train a simple, regression-based classifier (\eg~logistic regression or support vector classifier) that takes a pair of visualizations and predicts which one in the pair is preferred to the other.
The training data for the classifier is the matrix consisting of the \textit{frequency difference} vectors corresponding to $C'$.
A frequency difference vector, $d_x$ for a feature $x$ is computed as below:
\begin{equation}
    d_x = p_x - n_x,
    \label{eq:diff_vector}
\end{equation}
where $p_x$ is the frequency vector of the positive designs in the corpus for feature $x$, and $n_x$ is that for the negative designs (illustrated in \autoref{fig:feature-diff}).
When evaluating the trained classifier via cross validation, we obtain the average prediction.

To evaluate a candidate feature, we apply the following selection criteria (\autoref{fig:step7}).
\begin{enumerate}
    \item Improvement threshold: add a feature when it improves the performance by a certain degree (\eg~0.1\%).
    \item Variance threshold: even if a feature improves the average cross validation accuracy score, do not add the feature if the standard error of the cross validation accuracy scores is greater than this threshold. This can prevent adding features that are unlikely to generalize; yet if it is too low, the selection step may not be able to add any new feature.
\end{enumerate}

Without stopping criteria, the forward selection can result in too many iterations and explode the final feature set by adding features of low utility.
Thus, we apply the following halting criteria for forward selection.
\begin{enumerate}
    \item Pre-selection method lower bound: stop if a candidate feature's contribution estimate is below a certain value, as it is unlikely to offer generalizable design knowledge.
    \item Convergence: stop if the average cross validation accuracy score remains within a certain range (typically smaller than the improvement) over a certain number of consecutive times, as it is unlikely to provide any performance improvement.
    \item Not selected: stop if this process does not select any new feature consecutively for a certain number of times, due to the same reason as above.
    \item The size of selection: stop if the number of the selected features hits a certain threshold, as too many features may lead to overfitting.
\end{enumerate}
To summarize, these forward selection parameters are outlined in \autoref{tab:parameters}, along with values used in our evaluations.

\subsubsection{Step 8. Backward selection}

To see if there are any unnecessary features from the features selected in the previous step, we run backward selection.
This step trains a simple classifier by temporarily omitting a few features from the forward selection result.
If doing so improves the performance, using the same selection criteria, then we remove them from the feature selection set. 
The order of features to be tested is random.
Given $N$ features ($f_0, \dots, f_N)$ from the forward selection,
First, a single feature ($f_i$) is tested in a random order.
After testing at all single features ($N$ iterations), each pair of features ($(f_i, f_j)$ is tested in a random order ($\binom{N}{2}$ iterations).
We repeat this process by testing each \textit{n} features  ($\binom{N}{n}$ iterations) in a random order.
As this process can be computationally exhaustive, we include a parameter to limit the number of backward iterations. 

\subsection{Rendering}
\subsubsection{Step 9. Formulate logical rules}
The selected features in a key-value chain form do not directly support reasoning about visualization designs.
Because they are not in a formal logic format, it is not possible to trace different visualization elements.
For example, a combined key-value chain, \lstinline{scale_type.linear + data_type.continous}, needs to trace a scale and encoding definitions in a given chart spec that are matched based on their channel types (\eg~$x$).
We need to convert selected key-value chains into formal representations that a logic solver can use. 
Specifically, our rendering step returns features to Draco's \lstinline{entity} and \lstinline{attribute} representation format.

It is relatively straightforward to encode \lstinline{entity}s and \lstinline{attribute}s.
Given a key-value chain \lstinline{scale_type.linear}, for example, \lstinline{scale_type} is an attribute of a \lstinline{scale} entity.
We can render this as \lstinline{entity(scale,_,S)} (the type of entity) and \lstinline{attribute((scale,type),S,linear)} (the attribute's value).
Here, \lstinline{_} is a dummy parent marker because this feature does not trace its parent, and we use a variable (\lstinline{S}) for the identifier to apply this rule to arbitrary scale entities.

\begin{figure}
    \centering
    \includegraphics[width=\linewidth]{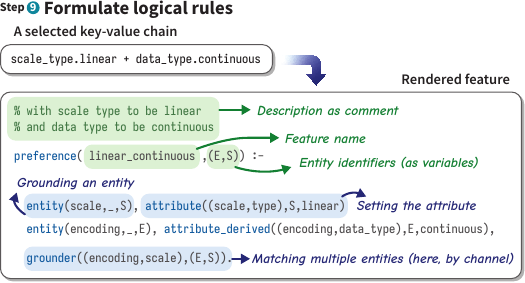}
    \caption{An example rendering of a key-value chain to a formal representation.}
    \label{fig:rendering}
    \Description{We render a key-value chain-based raw feature into a formal representation written using Draco expressions with a template-based description.}
\end{figure}

Features also require weight terms that proportionally promote or penalize that feature when reasoning about visualization designs. 
Following Draco-Learn~\cite{moritz2018:formalizing}, we train a simple, regression-based classifier to obtain coefficients that are then assigned as the feature weights. 
To enhance the understandability of features, we generate feature names and descriptions based on templates.
Combining these components, rendered output is illustrated in \autoref{fig:rendering}.

\section{Evaluation: Benchmark versus Draco 2}

To evaluate our synthesis method, we first conducted a benchmark comparison with Draco 2~\cite{yang2023:draco2} (Draco, hereafter), a knowledge base for common statistical visualizations.
We sought to assess both the generated feature set and prediction performance.
We believe this is a high bar for comparison, as Draco already provides 93-96\% of predication accuracy when using Draco-Learn~\cite{moritz2018:formalizing} method to determine the weights for Draco's manually authored soft constraints.

%tc:ignore
\begin{table}
    \caption{Data split for the benchmark and reusability evaluations. Gen(omics), H(old)-O(ut).}
    \label{tab:split}
    \Description{Data split for the evaluation studies. For cross validation, each Draco 2 Baseline split had 209 pairs, each Draco 2 Zeng+ aplit had 26 to 27 pairs, and each Genomic visualization split had 50 to 51 pairs. For holdout, Draco 2 Baseline had 184, Zeng+ had 23, and Genomic visualization had 44 pairs.}
    \centering
    \small
    \setlength\extrarowheight{2pt}
    \begin{tabular}{cc|ccccc|c|c} 
        \hline
        \multirow{2}{*}{\textbf{Exp.}} & \multirow{2}{*}{\textbf{Corpus}} & \multicolumn{5}{c|}{\textbf{Cross Validation}} & \multirow{2}{*}{\textbf{H.O.}} & \multirow{2}{*}{\textbf{Total}} \\ \cline{3-7}
        & & \textbf{1} & \textbf{2} & \textbf{3} & \textbf{4} & \textbf{5} &  & \\ \hline
        \multirow{3}{*}{\textbf{Draco2}} & \textbf{Baseline} & 209 & 209 & 209 & 209 & 209 & 184 & 1,229 \\
        & \textbf{Zeng+} & 27 & 27 & 26 & 26 & 26 & 23 & 155 \\ \cline{2-9}
        & \textbf{Total} & 236 & 236 & 235 & 235 & 235 & 207 & 1,384 \\ \hline \hline
        \textbf{Gen.} & \textbf{Total} & 51 & 51 & 50 & 50 & 50 & 44 & 296 \\ \hline
    \end{tabular}
\end{table}
%tc:endignore

\begin{figure}
    \centering
    \includegraphics[width=\linewidth]{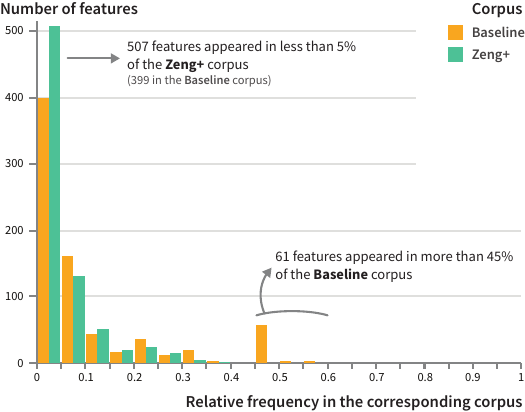}
    \caption{The ratio of features shared in each corpus. The \textbf{Baseline} corpus has 61 (out of 749) features appearing in more than 45\% of its pairs (1,229 pairs). In contrast, most features in the \textbf{Zeng+} set are shared in less than 30\% of its 155 pairs, with 507 features appearing in less than 5\% of them.}
    \label{fig:shared-features}
    \Description{The distribution of shared features across design pairs in the Baseline and Zeng-plus data sets. 61 features appeared more than 45\% of the Baseline corpus. 507 features appeared less than 5\% of the Zeng-plus corpus, while 399 features did so in the Baseline corpus.}
\end{figure}

\subsection{Experiment Setup}

\subsubsection{Design Corpus}
We used a corpus of 1,384 chart pairs collected from 30 graphical perception studies by Zeng~\ea~\cite{zeng2023:dataset,zeng2024:tooManyCooks}.
This corpus includes 1,229 pairs derived from Mackinlay's APT~\cite{mackinlay1986:APT} (67), Kim~\ea~\cite{kim2018:encodings} (1,152), and Saket~\ea~\cite{saket2018:encodings} (10) used in prior Draco studies (\textbf{Baseline}), as well as 155 pairs from the other 27 studies (\textbf{Zeng+}).
The \textbf{Baseline} set includes basic designs with overlapping design properties (\eg~comparing \textit{y} vs. size) across pairs.
On the other hand, those in the \textbf{Zeng+} set exhibit heterogeneous, specific features (\eg~time-series trend via line vs. heatmap~\cite{albers2014:task}), forming a sparse data set, where visualization pairs share less common design features.
Specifically, the pairs in the \textbf{Baseline} set shared 61 extracted unary features about half of the time (>45\%); while those in the \textbf{Zeng+} did not have such features (\autoref{fig:shared-features}). 
We took 15\% pairs as a holdout test set and divide the rest into five cross-validation splits for selection (see \autoref{tab:split}).
During the selection steps, we used one split for test and the rest for training.

\subsubsection{Configurations}
To gauge the utility of each pre-selection measure (Step 5), we use selection configurations that ablate each one of them as well as one that uses all of them (in total, seven ablations and one including all). 
We call each of them \textbf{selection process}.
To see how selection outcomes change, we use pipelines with only the \textbf{Baseline} or \textbf{Zeng+} datasets, in addition to one that includes all of the data.
In total, there are 24 configurations (3 datasets $\times$ 8 metric ablations). 
We use the selection parameters in \autoref{tab:parameters} (Benchmark column), which we determined by running several iterations to identify values that converge to selecting neither too few nor too many features.

\subsubsection{Providing A Priori Knowledge}
Draco encodes some critical knowledge that can not be determined from the individual properties in a design spec.
Examples of such information include the cardinality of a discrete encoding (which needs to consider binning, unique values, aggregation, \etc) and potential risks of overplotting (\eg~a plot with discrete \textit{x} and \textit{y} having no aggregate). 
We encoded these using our helper function, which in turn converts to logical expressions (see Step 2). 

\subsubsection{Performance Metric}
We primarily evaluate performance using prediction accuracy.
We use a simple, regression-based model that takes a pair of visualization designs (represented as the selected and candidate features at each step) and predicts which one of them should be preferred.
We can obtain prediction accuracy from this process, which we used as an internal metric for evaluating the performance. 
We assume a final knowledge base, which reasons about a design by summing corresponding weight terms, follows a similar behavior of regression-based linear models (specifically, support vector classifier with a linear kernel in our case).
This internal prediction accuracy serves as a proxy method for running the entire knowledge base given that the regression-based model and knowledge base both use a linear sum for the final judgment.
Running the entire knowledge base takes a lot longer time (up to a few minutes per iteration).
This way, we can speed up selection iterations, which number in the tens of thousands. 

\subsection{Results}

\begin{figure}
    \centering
    \includegraphics[width=\linewidth]{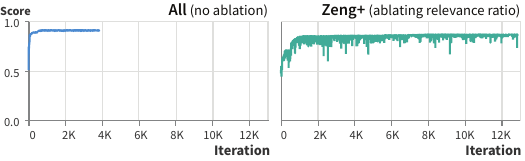}
    \caption{Forward selection iterations from our benchmark study. Forward selection starts with an initial set of features that are likely to contribute to performance, providing a high score jump in the beginning. Most of the time (16 out of 24 configurations), iterations terminated after about 3,000--7,000 iterations (left). Four selection processes terminated much later, after 10,000 iterations (right).}
    \label{fig:termination}
    \Description{Forward selection processes for the Draco benchmark evaluation. On the left, the selection process based on the entire data with no metric ablation terminated after about 4,000 iterations, while the selection process based on the Zeng-plus data with determinance ratio ablated terminated after 13,000 iterations.}
\end{figure}

\subsubsection{Termination}
Most forward selection processes (16/24) terminated after about 3,500 to 7,000 iterations (4,850 on average).
Another seven processes terminated between 7,000 and 13,000 iterations, among which five processes terminated after 10,000+ iterations.
Regarding how they halted, one process terminated after 2,661 iterations.
One process (\textbf{Baseline data} with difference count ablated) terminated after reaching the convergence threshold (200). 
The others terminated after not adding new features more times than the threshold (3,000).
Across all processes, the backend selection steps terminated after 10,000 iterations.

To understand the coverage of the selection processes, we break down the number of features considered at each step.
After Step 1--4 (extraction), we obtained 77,619 features, and in general, 6,000 to 40,000 features were filtered using the pre-selection metric threshold (at Step 5).
At Step 6 (forward), our selection processes explored 10 to 80\% of those filtered features.
For example, the selection process using the entire corpus without pre-selection metric ablation filtered in 47,803 candidate features and considered 3,833 (8\%) among them.
Other selection processes considered 8 to 50\% of those filtered features, with one exception (\textbf{Zeng+} with LDA score ablated; 6,745 filtered in and 5,892 (87\%) considered).

%tc:ignore
\begin{table}
    \caption{Prediction accuracy scores for the best performing synthesis selection processes compared to the Draco 2 feature set. The same cross validation and holdout splits apply across conditions.}
    \label{tab:perf1}
    \Description{The prediction accuracy scores for the best performing selection processes.}
    \centering
    \small
    \setlength\extrarowheight{2pt}
    \resizebox{\linewidth}{!}{
    \begin{tabular}{lllcc} 
        \hline
        \textbf{Test set} & \textbf{Corpus} & \textbf{Method} & \textbf{CV} & \textbf{Holdout} \\ \hline
        All      & All      & Draco 2               & 89.29 & 87.43 \\ 
                 &          & Synthesis: No Ablation                   & 90.91 & \textbf{89.85} \\ \cline{2-5}
                 & Baseline & Draco 2               & 88.99 & 86.47 \\ 
                 &          & Synthesis: No Specificity        & 89.42 & \textbf{87.20} \\ \cline{2-5}
                 & Zeng+    & Draco 2               & 63.80 & 70.05 \\ 
                 &          & Synthesis: No Appearance Count   & 88.23 & \textbf{85.51} \\ \hline \hline
        Baseline & All      & Draco 2               & 93.16 & \textbf{94.18} \\ 
                 &          & Synthesis: No Difference Count   & 94.35 & 93.53 \\ \cline{2-5}
                 & Baseline & Draco 2               & 92.92 & \textbf{93.82} \\ 
                 &          & Synthesis: No LDA Score          & 92.92 & 93.08 \\ \cline{2-5}
                 & Zeng +   & Draco 2               & 62.97 & 72.17 \\ 
                 &          & Synthesis: No Appearance Count   & 90.19 & \textbf{89.46} \\ \hline \hline
        Zeng+    & All      & Draco 2               & 66.38 & 72.58 \\ 
                 &          & Synthesis: No Determinance Ratio & 79.19 & \textbf{78.39} \\ \cline{2-5}
                 & Baseline & Draco 2               & 57.92 & 55.16 \\ 
                 &          & Synthesis: No Specificity        & 65.19 & \textbf{61.61} \\ \cline{2-5}
                 & Zeng+    & Draco 2               & 70.50 & 77.74 \\ 
                 &          & Synthesis: No Difference Count   & 78.10 & \textbf{81.94} \\ \hline
    \end{tabular}
    }
\end{table}
%tc:endignore

\subsubsection{Performance}
Compared to Draco's features, our method achieved slightly improved performance for \textbf{Baseline} data and significant improvements for \textbf{Zeng+} data (\autoref{tab:perf1}), demonstrating usefulness in a recommender context (\textbf{G1}).
When tested on the entire corpus, our method achieved a 1--15\% improvement in prediction accuracy on the holdout test set.
When tested on the \textbf{Baseline} corpus, our synthesis results are competitive (within 1\%) of Draco. Interestingly, when testing on the \textbf{Baseline} data but selecting on the \textbf{Zeng+} corpus, our models achieved a 17\% improvement over Draco, suggesting improved generalization.
When tested on the \textbf{Zeng+} corpus, our models improved performance by 4--6\%.

While \autoref{tab:perf1} shows selection processes without LDA score, there were cases where omitting LDA score achieved lowest performance.
As an ad-hoc analysis, we compared the holdout prediction accuracy by ablated metrics.
One-way ANOVA results did not find any significant influence of ablated metrics (F-statistic: $0.13$; p-value: $0.99$).
$\chi^2$ tests (for each test data) also did not find a significant effect of ablated metrics and selection corpus on prediction accuracy ($\chi^2$-statistic: $[0.0, 0.0, 0.0]$; p-value: $[1.0, 1.0, 1.0]$).

\newcommand{\smalldesc}[1]{{\Small #1}}
\newcommand{\featName}[1]{{\Small \texttt{#1}}}

%tc:ignore
\begin{table}
    \caption{Top 10 features selected by our synthesis method in the benchmark study. The direction of each feature is the sign of their average weights from the 24 selection processes. *The use of a linear scale has a lower average weight than that for the use of an ordinal scale, which indicates that using a linear scale is generally preferred to using an ordinal scale.}
    \label{tab:top20}
    \Description{Top 10 commonly selected features by our methods.}
    \centering
    \small
    \setlength\extrarowheight{2pt}
    \resizebox{241pt}{!}{%
    \begin{tabular}{rlcc} 
        \hline
        \textbf{Feature} & \textbf{Description} & \textbf{\#} & \textbf{Direction}  \\ \hline
        \featName{y\_interesting\_true}
            & \smalldesc{an interesting field on \textit{y}}	&	24 & Positive \\
        \featName{color} 
            & \smalldesc{having a color channel}            	&	22 & Positive \\
        \featName{ordinal}
            & \smalldesc{using an ordinal scale}	            &	22 & Negative \\
        \featName{n\_facet\_upper\_1}
            & \smalldesc{having no facet}           	        &	21 & Positive \\
        \featName{color\_interesting\_true} 
            & \smalldesc{an interesting field on color}	        &	20 & Negative \\
        \featName{size}	
            & \smalldesc{having a size channel}	                &	20 & Negative \\
        \featName{coordinates\_cartesian}
            & \smalldesc{using Cartesian coordinates}       &	19 & Positive \\
        \featName{x\_interesting\_true}
            & \smalldesc{an interesting field on \textit{x}}	&	19 & Positive \\
        \featName{linear}
            & \smalldesc{using an linear scale}              	&	19 & Negative* \\
        \hline
    \end{tabular}
    }
\end{table}

%tc:endignore

\subsubsection{Features Learned}\label{sec:bench:feat}
In general, our method selected 34 to 96 features (67.88 on average), much smaller than Draco's set of 147 features.
Specifically, selection processes using the entire corpus learned 43 to 96 features (81.5 on average), and those using the \textbf{Zeng+} subset learned 53 to 96 features (mean: 78.63).
On the other hand, the \textbf{Baseline} subset learned 34 to 58 features (43.5 on average).
Given that the \textbf{Zeng+} corpus represents a sparse pool of design features as shown above, it makes sense that the selection processes based on it learn more features.

Our synthesis method learned a different set of rules while sharing some core features as listed in \autoref{tab:top20}.
For encoding, features related to \textit{x}, \textit{y}, \textit{color} channels, and ordinal scale are most commonly learned as they are common design choices in basic visualization designs.
Our method also learned features that prefer \textit{interesting} fields (\ie~data fields of primary interest of a creator) to be mapped to position channels ($x$ and $y$), aligned with prior findings~\cite{mackinlay1986:APT,kim2018:encodings,saket2018:encodings}.
Furthermore, our method selected features in favor of non-faceted views and Cartesian coordinates.

We measured the cosine similarity between each pair of selection processes in terms of features selected and weight terms obtained.
Selection processes using the \textbf{Baseline} subset generally had cosine similarities of $0.2-0.7$, which makes sense given the \textbf{Baseline} designs share common design properties.
However, comparing selection processes based only on the \textbf{Baseline} and only on \textbf{Zeng+} had very low cosine similarities (below $0.25$) as they cover different design characteristics. 

To qualitatively assess the generalizability of the features synthesized using our method, we compared them to the Draco features.
For this analysis, we used the 76 synthesized features from a selection process with no ablation using the entire dataset.
For each selected feature, we looked at (1) whether the same feature appears in the Draco feature set; (2) whether it is a union of several features from Draco; and (3) whether it is a subset of a feature in Draco.
We also inspected those that have too narrow a scope as a potential cause of overfitting and lower holdout scores. 

First, 21 (out of 76) features exactly match a corresponding Draco feature, such as \lstinline{linear_x} (using a linear scale for $x$) and \lstinline{encoding_aggregate_mean} (having mean aggregate).
Next, 10 features are quite similar to those in Draco.
The \lstinline{summary_row_interesting_true} feature (having an ``interesting'' variable as a row facet for a summary task) from our method is similar to \lstinline{interesting_row} in Draco (no task distinction).
Twenty-one features are a union of individual features in Draco.
For example, Draco has \lstinline{value_discrete_color} and \lstinline{value_continuous_color} features for having a discrete or continuous color channel for a value task, while our method selects a \lstinline{value_color} feature that captures both. 
Similarly, there are 21 features that jointly capture different features in Draco.
For instance, our method selected \lstinline{point_x_data_type_continuous}: having a continuous $x$ channel for point mark.
This feature is a conjunction of \lstinline{*_point} (the use of point mark) and \lstinline{continuous_x} (continuous $x$). 

Lastly, our method did not capture 79 Draco features (out of 147) mostly because our dataset did not cover those features (\eg~binning; aggregation methods other than mean; log scale).
At the same time, our method did not select many of Draco's ``no overlap'' features.
We attribute this to the fact that most of the charts were ``well-designed'' with fewer of those concerns. 
Ten of the selected features were not fully addressed by Draco.
Those features appear to be redundant (\eg~having a mark) with zero weight term or relevant to data cardinality.
For example, \lstinline{color_cardinality_upper_15} captures using a color channel with less than 15 different color values\footnote{\lstinline{upper} means upper bound, not greater.}.
Our method likely selected those cardinality-related features because \textbf{Zeng+} pairs include cardinality-related features (\eg~using the same design but with different data sizes).

%tc:ignore
\begin{table}
    \caption{Using synthesis to augment an existing knowledge base: prediction accuracy scores for ad-hoc selection processes in which we used the Draco 2 features as the initial feature set. The same cross validation and holdout splits apply across conditions. Holdout accuracy values higher than those from the main results (\autoref{tab:perf1}) are boldfaced; while competitive ones ($\Delta <1\%$) are italicized.}
    \label{tab:perf2}
    \Description{The prediction accuracy scores for the ad-hoc selection processes, which had comparable performances to the original Draco in general.}
    \centering
    \small
    \setlength\extrarowheight{2pt}
    \begin{tabular}{llcc} 
        \hline
        \textbf{Test set} & \textbf{Corpus} & \textbf{CV} & \textbf{Holdout} \\ \hline
        All      & All      & 91.21 & 87.92 \\ 
                 & Baseline & 90.44 & \textit{86.96} \\ 
                 & Zeng+    & 84.58 & \textit{84.54} \\ \hline \hline
        Baseline & All      & 93.73 & \textit{94.14} \\ 
                 & Baseline & 94.26 & \textbf{94.30} \\ 
                 & Zeng +   & 85.93 & 86.74 \\ \hline \hline
        Zeng+    & All      & 71.24 & 76.13 \\ 
                 & Baseline & 60.17 & 53.87 \\ 
                 & Zeng+    & 73.96 & \textit{81.94} \\  \hline
    \end{tabular}
\end{table}
%tc:endignore

\subsubsection{Ad-hoc Analysis: Selection on Top of Draco Features}
Given a scenario where the goal is to examine and extend an existing knowledge base, we ran three ad-hoc selection processes.
In these selection processes, we used the existing Draco features as the initial selected features.
The forward selection step then adds each candidate feature in the order sorted by the pre-selection metrics.
The backward selection step also considers removing Draco features.
We did not ablate any pre-selection metrics for this scenario.

As shown in \autoref{tab:perf2}, the performance of the ad-hoc selection processes are comparable to the main results. 
However, prediction accuracy of the selection process based on \textbf{Zeng+} dropped when tested on the \textbf{Baseline} designs (\textbf{Zeng+} $\rightarrow$ \textbf{Baseline}). 
Accordingly, the features and learned weight terms did not generalize as well to the holdout test set. 

The selection process using the entire corpus selected eight features from the extracted candidates and removed four from the Draco features. 
Among the added ones, three of them were redundant (\lstinline{linear_scale}, \lstinline{y_interesting_true}, and \lstinline{color_interesting_true}, with positively directed weights).
Another three added features were special cases of the existing Draco features (\eg~\lstinline{area_mark} + \lstinline{continous_color}, with a negatively directed weight).
The other two features were related to data characteristics.
For example, the \lstinline{x_unique_under_70} (an $x$ channel with less than 70 unique values) had a negatively directed weight, penalizing encoding a non-quantitative field. 
We speculate that the four features removed from Draco were because of their low or inconsistent frequencies across the pairs.
For example, \lstinline{col_c_d} (using a column facet with both continuous and discrete position---$x$ and $y$---encodings) did not appear in our corpus.
On the other hand, \lstinline{summary_rect} (using a rectangle mark for a summary task) appears in both more preferred and less preferred designs in the corpus.
This information could motivate collecting more data or designing an experiment that can specifically target those features.

Interestingly, the ad-hoc selection process based on the \textbf{Zeng+} corpus selected 55 new features and removed 9 from Draco. 
Those removed features were more generic features (\eg~\lstinline{ordinal_scale}, \lstinline{summary_rect}).
In contrast, those added ones tended to be more specific, such as encoding a discrete field of interest using a row facet (with a positively directed weight) and using a color channel for bar mark (with a negatively directed weight). 
\section{Evaluation: Synthesizing a Genomics Visualization Knowledge Base}
\label{sec:eval:genomics}

To test the applicability of our method to other visualization domains, we applied them to genomics data visualization. 
One of the key characteristics of genomics data visualizations is that they usually consist of multiple views for comparison across different parts of genomic data~\cite{l2022:genomics}.
This distinction from single-view visualizations allows us to assess how well our method synthesizes cross-view features. 
Specifically, we used genomics visualizations expressed using Gosling~\cite{lyi2022:gosling}, a declarative domain-specific language for interactive genomics data visualizations.

\begin{figure*}
    \centering
    \includegraphics[width=\textwidth]{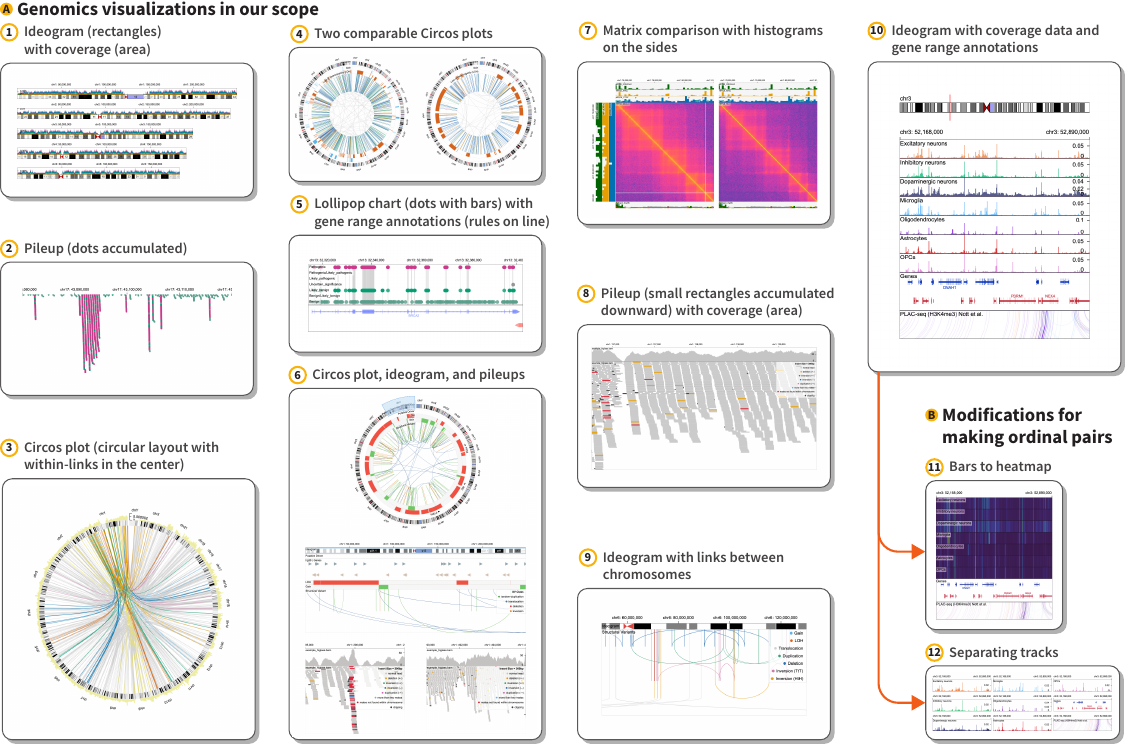}
    \caption{(A) Genomics visualizations used as input data for synthesis of a new knowledge base. (B) Example design modifications used to create visualization variants that are then grouped into design pairs.}
    \label{fig:genomic-charts}
    \Description{Genomics visualization designs used in our evaluation. They include: (1) ideogram with coverage information, (2) pile-up chart, (3) a single Circos plot, (4) two horizontally arranged Circos plot, (5) a lollipop chart, (6) a mixture of Circos, ideogram, and pile-ups, (7) matrix comparison with histogram, (8) pile-up with coverage information, (9) ideogram with between-links, and (10) ideogram with coverage and gene range views. These charts are slightly modified to produce labeled visualization pairs. For example, bars are changed into heatmaps (11) and tracks are separated into different charts (12).}
\end{figure*}

\subsection{Experiment Setup}

\subsubsection{Design Corpus}
While there are archives for genomics visualizations~\cite{nusrat2019:genomic,agv,genocat}, they are not prepared as design pairs. 
Thus, we curated a new corpus based on representative examples from the Gosling gallery,\footnote{\url{https://gosling-lang.org/examples/}} inspired by prior work on data augmentation techniques for visualization~\cite{kim2025:augmentation}.
We began with ten genomics visualizations, covering comparative matrices, genome sequence, Circos plot~\cite{krzywinski2009:circos} (circular chart with links between elements), annotated genes, and so on (\autoref{fig:genomic-charts}-A).
These charts are widely adopted in genomics research according to a prior survey~\cite{nusrat2019:genomic}.
For each initial visualization $V_0$, we permute design elements to generate variants $V_1, ..., V_n$, grouped them into pairs, $(V_0,V_1), (V_0,V_2), \ldots, (V_{n-1},V_n)$, and had a genomics visualization expert label those pairs (illustrated in \autoref{fig:genomic-charts}-B).
Specifically, we gradually applied transformations that degrade the original designs (\eg~removing interaction for a multi-view chart, applying layouts that are unusual for given data structures), which we include in the Supplementary Material.
This process resulted in a corpus of 296 pairs, which we divided into a holdout set (15\%) and five cross-validation splits, summarized in \autoref{tab:split}.
This experiment is intended as a proof of concept; we do not claim that this provides a comprehensive corpus. 

\subsubsection{Configurations and Performance Metric}
Given that we have a single, relatively small corpus, we do not have a subset corpus, yet we use the same pre-selection metric ablation as before (\ie~total of eight configurations). 
We also use the same prediction accuracy metric.
As we do not have an existing system to directly compare to, we qualitatively assess the final selected features with respect to how we modified the original designs to enumerate pairs in order to see whether the features capture those modifications.

\subsubsection{Providing A Priori Knowledge}
Gosling specifies genomics visualizations at a low level for some of the complex chart types, similar to how Vega~\cite{satyanarayan:vega2016} specifies a chart design. 
Thus, extracting features from a raw specification itself may not provide useful information by looking at low-level features that make sense in a bigger picture.
Therefore, we inferred chart types that are widely adopted by genomics researchers.
For example, we can characterize chromosome ideograms (\autoref{fig:genomic-charts}-A1, 6, 9, \& 10) as an overlay of rectangles and two triangles. 
Similarly, a Circos plot~\cite{krzywinski2009:circos} (\autoref{fig:genomic-charts}-A2) can be characterized as a circular format with links between elements around the circle edge.
Another important aspect is the location of views to consider multiple-view constraints (\cf~\cite{qu2018:multiview}), such as having shared \textit{x} and \textit{y} axes. 
We encoded inference rules for a priori knowledge, including representative chart types and \textit{x} and \textit{y} collocations, using our helper function.
We provide a detailed list of such inference rules in the Supplementary Material.

\subsection{Results}

\subsubsection{Termination}
All eight selection processes terminated after about 1,200 to 5,700 iterations (4,310 on average).
After Step 4, we obtained 7,883 features, and in general, about 4,500 to 5,500 features were filtered using the pre-selection metric threshold.
Our selection processes terminated after looking at 85\% (one) to 95-100\% (seven) of those filtered in (97.5\% on average).

\subsubsection{Performance}
Our selection processes resulted in prediction accuracy ranging from 94.64\% (relevance ablated) to 98.40\% (determinance ratio ablated) in cross-validation.
On the holdout test set, the prediction accuracy ranged from 82.95\% (relevance ablated) to 97.72\% (determinance ratio ablated).
Prediction accuracy consistently dropped by 0.7--11.7\% from cross validation to holdout set, presumably because of the small corpus. 
The number of selected features did not have a strong correlation to the amount of performance drop (correlation: $-0.1$).

\begin{figure}
    \centering
    \includegraphics[width=\linewidth]{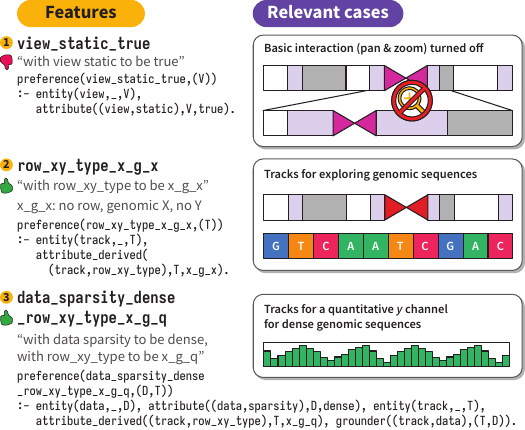}
    \caption{Features selected for genomics visualization.}
    \label{fig:gen_feat}
    \Description{Our method selected features for view interactivity, row, x, and y layout, and a combination of data sparsity and layout.}
\end{figure}

\subsubsection{Features Selected}
In addition to generic features like the number of genomic variables or the bounds for unique values of a categorical variable, our selection processes learned genomic visualization-specific features (24--49 features in total; average: 36.5). 
As illustrated in \autoref{fig:gen_feat}, we describe a few examples from the features selected by the best performing process (determinance ratio ablated), and the entire features are available in the Supplementary Material.
For example, genomics visualization tends to display a massive amount of data, so being able to make simple interactions like panning and zooming is crucial, for which our method selected a feature that penalizes a static view (\lstinline{view_static_true}).
Genomics visualizations often include a view that shows a genome sequence, which our method captured as \lstinline{row_xy_type_x_g_x}.
Here, \lstinline{x_g_x} means that a chart (track) has no row and \textit{y} channels, but has the \textit{x} channel for a genomic variable. 
The last example feature considers data characteristics. 
When a dataset is densely distributed, a common design choice is to plot their quantitative distribution  (\eg~\autoref{fig:genomic-charts}-10).
Our method selected a feature that suggests using a quantitative \textit{y} channel, as shown in \autoref{fig:gen_feat}-3.

\subsubsection{Expert feedback}

To assess how helpful the selected features are, we invited two experts (A and B) who frequently visualize and analyze genomic data.
Both experts held a Ph.D. degree with research focus on genomic visualizations. 
We provided them with the list of features (determinance ratio ablated) and asked them to assess how meaningful the selected features are in practice.
We prompted them with ``How helpful is this feature for reasoning about whether a given genomic visualization is useful or not?'' and they responded on a four-point scale:
\begin{itemize}
    \item Irrelevant (0): this feature has nothing to do with determining design effectiveness.
    \item Not so helpful (1): this feature could potentially be useful, but I cannot think of such cases at the moment.
    \item Helpful (2): I can see how this feature is relevant to many cases.
    \item Very helpful (3): this feature always needs to be considered in genomics visualizations.
\end{itemize}
We described features verbally and with text to not restrict experts' reasoning.
Yet, we provided examples when they had difficulty coming up with relevant design cases immediately.
We concluded each session by asking for aspects that our features may be missing.

\begin{figure}
    \centering
    \includegraphics[width=\linewidth]{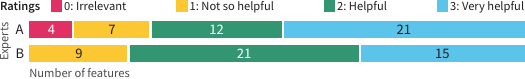}
    \caption{Feature ratings per expert reviewer.}
    \label{fig:feedback-result}
    \Description{Experts generally rated our features to be helpful. Expert A rated four irrelevant, seven not so helpful, twelve helpful, and twenty-one very helpful. Expert B rated zero irrelevant, nine not so helpful, twenty-one helpful, and fifteen very helpful.}
\end{figure}

Overall, both experts found the selected features to be helpful (75-80\%) in reasoning about genomic visualization designs, as illustrated in \autoref{fig:feedback-result}.
Features that both experts found to be Very helpful include the use of color channel (\lstinline{channel_color}), the use of bar and rectangle marks (\lstinline{mark_bar} and \lstinline{mark_rect}), having a static/interactive view (\lstinline{view_static_true}), and coordinated \textit{x} axes.
On the other hand, they found overly specific features like the use of color channel for line marks and flipping the \textit{y} axis for a circular layout (\autoref{fig:genomic-charts}-2, in a circular mode) to be Not so helpful or Irrelevant.

The experts agreed on 13 out of 45 features, and most disagreements (24) had differences of one level: between Helpful and Very Helpful (15); between Helpful and Not so helpful (8); and between Not so helpful and Irrelevant (1).
For example, Expert B found a generic feature for a flipped \textit{y} axis to be Helpful but did not see its benefit for a circular layout (Not so helpful). 
For the same features, Expert A rated Not so helpful and Irrelevant, respectively, because they generally did not prefer the pileup design.
Their ratings greatly disagreed on a feature for a circular chart with a clearly defined genomic region along the \textit{x} axis.
Expert A found it to be Irrelevant because it does not affect the effectiveness of a design per se while Expert B rated it Very helpful because the genomic region is highly relevant information.

The experts also commented on the features lacking in our knowledge base, implying needs for curating more relevant design pairs.
For example, both wished to see features relating to brushing interaction (focus+context) given its importance in exploring large genomic regions. 
Expert A mentioned the absence of triangle marks as they are widely used to express directionality of genomic information (\eg~plus and minus strands).
Expert B pointed out potential features regarding file formats because visualization designs sometimes depend on genomic data file formats (\eg~BigWig, BED).
Albeit done with a small pool, expert feedback show the potential applicability of our method to different visualization domains while suggesting considerations for future work.

\begin{figure}
    \centering
    \includegraphics[width=\linewidth]{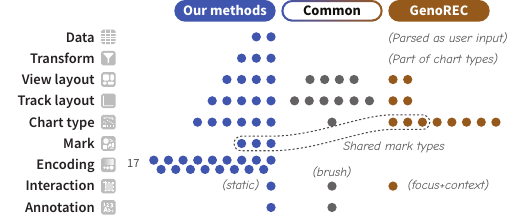}
    \caption{Comparing features selected using our method (determinance ratio ablated) and GenoREC~\cite{pandey2023:genorec}'s decision components. When a selected feature is about multiple categories (\eg~\autoref{fig:gen_feat}-3), they appear for all of them. The original GenoREC paper terms ``chart type'' as ``encoding.'' Here, we use encoding to refer strictly to mappings between data and visual channels.}
    \label{fig:gen_cmp}
    \Description{When compared to GenoREC, our method selects more encoding-related features while GenoREC considers mark types and encodings together as chart types. Both methods shared some common features regarding view and track layouts.}
\end{figure}

\subsubsection{Comparing Features to GenoREC}
GenoREC~\cite{pandey2023:genorec} is a knowledge-driven recommendation model that has top-down decision steps with respect to encoding, layering, within-view layout, data segmentation, between-view layout, and interactivity.
GenoREC's decision matrices are manually defined by referring to domain knowledge, such as prior work and expert discussion.
Given input genomics data, for example, GenoREC first filters appropriate encodings and marks. Then, it selects several layout options (\eg~stack vs. overlay and linear vs. circular). Lastly, GenoREC considers interactivity (coordinated views and focus+context).
While GenoREC's approach is sequential for each design aspect, our goal is to produce a knowledge base that reason about design features for different aspects simultaneously.
To reason about how our approach expands a prior understanding of genomics visualization design, as well as limitations, we qualitatively compared features selected from our method to the components of GenoREC's decision steps. 

As plotted in \autoref{fig:gen_cmp},
our work and GenoREC commonly captured all high-level design categories defined in the visualization taxonomy for genomics data~\cite{nusrat2019:genomic}.
Our features consider mark types and encoding channels separately (\eg~bar, \textit{y}) and together as chart types (\eg~heatmap), while GenoREC is designed to consider them together.
Our work exclusively includes features related to more advanced encoding (semantic zooming) and data characteristics (data cardinality and data transformation).
Given the scale of genomics data (\eg~3 billion bases for human genomes), using semantic zooming that scales across multiple zoom levels is an important design property, as emphasized in the original Gosling paper~\cite{lyi2022:gosling}.
Our method captured how semantic zooming can be defined in Gosling and in which cases (\eg~specific data types).
Our method also selected a feature for the use of encoding channels contributing to visual clutter (\eg~stroke) when the data is dense.
On the other hand, our method did not capture features relating to user tasks as our dataset did not include relevant information.
For example, by providing an ``explore'' task, GenoREC can produce charts that show the entire genome instead of displaying a narrow genomic region. 
Similarly, a ``compare'' task in GenoREC produces side-by-side views for comparing multiple genomic regions.

\section{Discussion}

To automatically construct visualization design knowledge bases from scratch, we proposed a method for extracting candidate features from a design corpus, making a small selection from them, and rendering the selected features into formal representations for further use. 
Through a benchmark evaluation with Draco 2, our method selected informative and sensible features with high performance in predicting preferred visualization designs, meeting or exceeding the performance of existing knowledge bases with hand-authored features.
In addition, our application study for genomics visualization demonstrates that our method is feasible for visualization designs beyond basic chart types.
Below, we discuss future research directions to leverage our approaches and implications for visualization research. 

\bpstart{Who and how to apply our method}
We intend our synthesis method for system developers and researchers to build knowledge bases and recommenders, supporting easier creation of domain-specific visualizations.
Reflecting on our development procedures, we envision the following scenarios relevant for visualization community.
First, researchers could use them to assess the completeness of their corpus by reflecting on some initial selection results.
For instance, while building the corpus of genomics data, we had a conversation that data sparsity could impact design decisions.
Based on that, we included visualization pairs that can show such differences. 
Future work could further extend the corpus based on the expert feedback we solicited.
In the benchmark experiment (\autoref{sec:bench:feat}), similarly, there were some cases where our method selected data type-agnostic features while Draco has distinctions (\eg~\lstinline{value_color} vs. \lstinline{value_discrete/continuous_color}). 
This may indicate the needs for collecting more relevant designs given the potential importance of such distinctions.
Next, another use case could be to complement an existing knowledge base with novel features.
For example, as we ran the ad-hoc selection processes on top of Draco's design features as an initial selection, our method can function as a way to examine an existing knowledge base and find gaps in it.

\bpstart{Further application areas}
Our work looked at general-purpose visualizations for benchmarking and genomic data visualizations for testing applicability.
Our experiment results and the expert feedback on genomic visualization features demonstrate the feasibility of our method in other visualization domains (\textbf{G2}: Adoptability). 
We envision that future work could apply our method to spread knowledge-driven visualization authoring practices to other domains, such as network, uncertainty, and dashboard visualization.
Preparing a labeled design corpus will require major effort, as it is common across many ML-based methods.

Suppose a case where a developer is trying to build an uncertainty visualization design knowledge base to support statisticians.
First, the developer would need to curate a design corpus consisting of ordered pairs.
A key consideration in preparing a design corpus is to have multiple pairs that precisely target each design aspect.
Pairs looking at multiple design aspects could result in learning counterintuitive weight terms, and too few pairs for a design aspect could fail to select important features due to overfitting. 
For example, pairs for \textit{y}-axis scaling for confidence intervals (inspired by Hoffman~\ea~\cite{hofman2020:cipi}) would need to keep other design aspects (\eg~baseline mark type) as consistent as possible.
The next step is to express those designs using Draco expressions directly (like our benchmark corpus) or by building a parser (as we did for genomics visualizations).
The JSON-to-Draco parser\footnote{\url{https://dig.cmu.edu/draco2/facts/examples.html}} could be an option.
Then, the developer can start with feeding the corpus expressed in Draco to the implementation of our method (provided in the Supplementary Material). 
The final outcome could be much more meaningful if integrated with data analytic tools like Dziban~\cite{lin2020:dziban} and EVM~\cite{kale2024:evm}.
For complex visualization domains like dashboards, multiple knowledge bases could be useful by modularizing design features for individual charts and their compositions.
Lastly, while our benchmark considers two generic tasks (summary and value), future work could adopt more sophisticated visualization task or insight models (\eg~Pyxis~\cite{battle2024:pyxis}), potentially with natural language support (\eg~NL4DV~\cite{narechania2021:nl4dv}).

\bpstart{Extensions to other science areas}
Some science domains, such as audio perception and mass communication studies, rely on perceptual experiments.
On the one hand, psychological perception studies often test which method best helps people to perceive some signal.
For instance, audio perception studies for effective data sonification designs (\eg~Walker~\ea~\cite{walker2002:magnitude,walker2007:consistency}) assess different audio signals.
We could express such sonification designs using Erie~\cite{kim:erie2023}, a declarative grammar that intentionally resembles the Vega-Lite grammar~\cite{satyanarayan:vega-lite2017}, making it easy to convert sonification specifications to Draco expressions.
A common goal among those areas is to understand what properties are desirable. 
Thus, future research could apply our method by forming a corpus of comparative design pairs as outlined above.

\bpstart{Limitations}
Our work uses a corpus from prior work (benchmark) and one that a single expert curated (genomics). 
While these corpora could be relevant for scenarios where a single practitioner develops a knowledge base, they may not include cases from more recent studies that could have an impact on feature selection. 
Yet, our core contribution is the method to make it easy to extend and synthesize visualization knowledge bases. 
Next, we designed our method considering visualization design recommendation as an application given its value as a shared goal across visualization research.
Future work could take a different application scenario, such as visualization linting~\cite{chen2022:vizlinter,mcnutt2025:colorlinter} or profiling visualization corpora~\cite{srinivasan2025:dashboard,heer2010:zoo,zeng2024:tooManyCooks}, to extend our work.

\bpstart{Concluding remarks}
As an approach to distilling formal representations from ranked visualization design pairs, our work motivates an important next step: how to expand a visualization design knowledge pool using individual designs. 
Leveraging our approach along with prior work~\cite{kim2025:augmentation,yang2023:draco2,zeng2024:tooManyCooks,moritz2018:formalizing,hu19:vizml,li2022:structure,wu21:lq2}, we envision next-generation visualization knowledge bases that not only reason about effective designs, but also illuminate what is missing in our current understanding of the visualization design space.

%% The acknowledgments
\begin{acks}
This research was supported in part by NSF IIS Award (2402718).
\end{acks}

%tc:ignore
\section*{Supplementary Material}
The Supplementary Material (\url{https://osf.io/q7xmu}) provides the source code, raw data, results, and analysis scripts of our evaluation experiments and genomics visualization corpus.
%tc:endignore

\bibliographystyle{ACM-Reference-Format}
\bibliography{references}

@article{casner1991:boz,
	title        = {Task-analytic approach to the automated design of graphic presentations},
	author       = {Casner, Stephen M.},
	year         = 1991,
	journal      = {ACM Trans. Graph.},
	publisher    = {ACM},
	volume       = 10,
	number       = 2,
	pages        = {111–151},
	doi          = {10.1145/108360.108361}
}

@inproceedings{shahapure:silhouette,
  title={Cluster quality analysis using silhouette score},
  author={Shahapure, Ketan Rajshekhar and Nicholas, Charles},
  booktitle={2020 IEEE 7th international conference on data science and advanced analytics (DSAA)},
  pages={747--748},
  year={2020},
  organization={IEEE}
}

@article{vanderplas:altair,
  title={Altair: interactive statistical visualizations for Python},
  author={VanderPlas, Jacob and Granger, Brian and Heer, Jeffrey and Moritz, Dominik and Wongsuphasawat, Kanit and Satyanarayan, Arvind and Lees, Eitan and Timofeev, Ilia and Welsh, Ben and Sievert, Scott},
  journal={Journal of open source software},
  volume={3},
  number={32},
  pages={1057},
  year={2018},
  publisher={The Open Journal}
}

@article{hunter:matplotlib,
  Author    = {Hunter, J. D.},
  Title     = {Matplotlib: A 2D graphics environment},
  Journal   = {Computing in Science \& Engineering},
  Volume    = {9},
  Number    = {3},
  Pages     = {90--95},
  publisher = {IEEE COMPUTER SOC},
  doi       = {10.1109/MCSE.2007.55},
  year      = 2007
}

@article{krzywinski2009:circos,
  title={Circos: an information aesthetic for comparative genomics},
  author={Krzywinski, Martin and Schein, Jacqueline and Birol, Inanc and Connors, Joseph and Gascoyne, Randy and Horsman, Doug and Jones, Steven J and Marra, Marco A},
  journal={Genome research},
  volume={19},
  number={9},
  pages={1639--1645},
  year={2009},
  publisher={Cold Spring Harbor Lab},
doi={10.1101/gr.092759.109}
}

@article{l2022:genomics,
  title={Multi-view design patterns and responsive visualization for genomics data},
  author={L'Yi, Sehi and Gehlenborg, Nils},
  journal={IEEE transactions on visualization and computer graphics},
  volume={29},
  number={1},
  pages={559--569},
  year={2022},
  publisher={IEEE},
doi={10.1109/TVCG.2022.3209398}
}

@article{mittal1998:sage,
	title        = {Describing complex charts in natural language: a caption generation system},
	author       = {Mittal, Vibhu O. and Carenini, Giuseppe and Moore, Johanna D. and Roth, Steven},
	year         = 1998,
	journal      = {Comput. Linguist.},
	publisher    = {MIT Press},
	volume       = 24,
	number       = 3,
	pages        = {431–467}
}

@article{roth1988:sage1,
	title        = {Graphics and Natural Language as Components of Automatic Explanation},
	author       = {Roth, Steven F. and Mattis, Joe and Mesnar, Xavier},
	year         = 1988,
	journal      = {SIGCHI Bull.},
	publisher    = {ACM},
	volume       = 20,
	number       = 1,
	pages        = 76,
	doi          = {10.1145/49103.1046410}
}

@inproceedings{schroeder1992:visage,
	title        = {{VISAGE: an object-oriented scientific visualization system}},
	author       = {Schroeder, W.J. and Lorensen, W.E. and Montanaro, G.D. and Volpe, C.R.},
	year         = 1992,
	month        = {Oct},
	booktitle    = {Proceedings Visualization '92},
	publisher    = {IEEE},
	pages        = {219--226},
	doi          = {10.1109/VISUAL.1992.235205}
}

@article{moritz2018:formalizing,
	title        = {Formalizing visualization design knowledge as constraints: Actionable and extensible models in Draco},
	author       = {Moritz, Dominik and Wang, Chenglong and Nelson, Greg L and Lin, Halden and Smith, Adam M and Howe, Bill and Heer, Jeffrey},
	year         = 2018,
	journal      = {IEEE Trans. Vis. Comput. Graph.},
	publisher    = {IEEE},
	volume       = 25,
	number       = 1,
	pages        = {438--448},
	doi          = {10.1109/TVCG.2018.2865240}
}

@article{mackinlay1986:APT,
	title        = {Automating the design of graphical presentations of relational information},
	author       = {Mackinlay, Jock},
	year         = 1986,
	journal      = {ACM Trans. Graph.},
	publisher    = {ACM},
	volume       = 5,
	number       = 2,
	pages        = {110--141},
	doi          = {10.1145/22949.22950}
}

@inproceedings{yang2023:draco2,
	title        = {Draco 2: An Extensible Platform to Model Visualization Design},
	author       = {Yang, Junran and Gyarmati, P\'{e}ter Ferenc and Zeng, Zehua and Moritz, Dominik},
	year         = 2023,
	booktitle    = {IEEE Visualization and Visual Analytics},
	publisher    = {IEEE},
	series       = {VIS '23},
	pages        = {166--170},
	doi          = {10.1109/VIS54172.2023.00042}
}

@article{schmidt2024:dracova,
	title        = {Visual Analytics for Understanding Draco’s Knowledge Base},
	author       = {Schmidt, Johanna and Pointner, Bernhard and Miksch, Silvia},
	year         = 2024,
	journal      = {IEEE Trans. Vis. Comput. Graph.},
	publisher    = {IEEE},
	volume       = 30,
	number       = 1,
	pages        = {392--402},
	doi          = {10.1109/TVCG.2023.3326912}
}

@article{zeng2024:tooManyCooks,
	title        = {Too Many Cooks: Exploring How Graphical Perception Studies Influence Visualization Recommendations in Draco},
	author       = {Zeng, Zehua and Yang, Junran and Moritz, Dominik and Heer, Jeffrey and Battle, Leilani},
	year         = 2024,
	journal      = {IEEE Trans. Vis. Comput. Graph.},
	publisher    = {IEEE},
	volume       = 30,
	number       = 1,
	pages        = {1063--1073},
	doi          = {10.1109/TVCG.2023.3326527}
}

@inproceedings{wongsuphasawat2016:compassQL,
	title        = {Towards a general-purpose query language for visualization recommendation},
	author       = {Wongsuphasawat, Kanit and Moritz, Dominik and Anand, Anushka and Mackinlay, Jock and Howe, Bill and Heer, Jeffrey},
	year         = 2016,
	booktitle    = {Workshop on Human-In-the-Loop Data Analytics},
	publisher    = {ACM},
	series       = {HILDA '16},
	doi          = {10.1145/2939502.2939506}
}

@inproceedings{lifschitz2008:asp,
	title        = {What Is Answer Set Programming?},
	author       = {Lifschitz, Vladimir},
	year         = 2008,
	booktitle    = {AAAI Conference on Artificial Intelligence},
	publisher    = {AAAI},
	series       = {AAAI '23},
	pages        = {1594--1597}
}

@article{wongsuphasawat2016:voyager,
	title        = {Voyager: Exploratory Analysis via Faceted Browsing of Visualization Recommendations},
	author       = {Wongsuphasawat, Kanit and Moritz, Dominik and Anand, Anushka and Mackinlay, Jock and Howe, Bill and Heer, Jeffrey},
	year         = 2016,
	journal      = {IEEE Trans. Vis. Comput. Graph.},
	volume       = 22,
	number       = 1,
	pages        = {649--658},
	doi          = {10.1109/TVCG.2015.2467191}
}

@inproceedings{wongsuphasawat2017:voyager,
	title        = {Voyager 2: Augmenting visual analysis with partial view specifications},
	author       = {Wongsuphasawat, Kanit and Qu, Zening and Moritz, Dominik and Chang, Riley and Ouk, Felix and Anand, Anushka and Mackinlay, Jock and Howe, Bill and Heer, Jeffrey},
	year         = 2017,
	journal      = {ACM Human Factors in Computing Systems},
	booktitle    = {Proc. CHI},
	publisher    = {ACM},
	series       = {CHI '17},
	pages        = {2648--2659},
	doi          = {10.1145/3025453.3025768},
}

@article{kim2023:dupo,
	title        = {Dupo: A Mixed-initiative Authoring Tool for Responsive Visualization},
	author       = {Kim, Hyeok and Rossi, Ryan and Hullman, Jessica and Hoffswell, Jane},
	year         = 2024,
	journal      = {IEEE Trans. Vis. Comput. Graph.},
	volume       = 30,
	number       = 1,
	pages        = {934--943},
	doi          = {10.1109/TVCG.2023.3326583}
}

@article{gebser2014:clingo,
	title        = {Clingo  {ASP} + Control: Preliminary Report},
	author       = {Martin Gebser and Roland Kaminski and Benjamin Kaufmann and Torsten Schaub},
	year         = 2014,
	journal      = {International Conference on Logic Programming},
	note         = {\url{https://arxiv.org/abs/1405.3694}},
	archiveprefix = {arXiv},
	eprint       = {1405.3694},
	eprinttype   = {arxiv}
}

@article{senay1994:vista,
	title        = {A knowledge-based system for visualization design},
	author       = {Senay, H. and Ignatius, E.},
	year         = 1994,
	journal      = {IEEE Comput. Graph. Appl.},
	volume       = 14,
	number       = 6,
	pages        = {36--47},
	doi          = {10.1109/38.329093}
}

@article{wang2025:dracogpt,
	title        = {{DracoGPT}: Extracting Visualization Design Preferences from Large Language Models},
	author       = {Wang, Huichen Will and Gordon, Mitchell and Battle, Leilani and Heer, Jeffrey},
	year         = 2025,
	journal      = {IEEE Trans. Vis. Comput. Graph.},
	volume       = 31,
	number       = 1,
	pages        = {710--720},
	doi          = {10.1109/TVCG.2024.3456350}
}

@inproceedings{zeng2023:dataset,
	title        = {A Review and Collation of Graphical Perception Knowledge for Visualization Recommendation},
	author       = {Zeng, Zehua and Battle, Leilani},
	year         = 2023,
	booktitle    = {ACM Human Factors in Computing Systems},
	publisher    = {ACM},
	address      = {New York, NY, USA},
	series       = {CHI '23},
    pages        = {820:1--820:16},
	doi          = {10.1145/3544548.3581349}
}

@article{saket2018:encodings,
	title        = {Evaluating Interactive Graphical Encodings for Data Visualization},
	author       = {Saket, Bahador and Srinivasan, Arjun and Ragan, Eric D. and Endert, Alex},
	year         = 2018,
	journal      = {IEEE Trans. Vis. Comput. Graph.},
	volume       = 24,
	number       = 3,
	pages        = {1316--1330},
	doi          = {10.1109/TVCG.2017.2680452}
}

@article{kim2018:encodings,
	title        = {Assessing Effects of Task and Data Distribution on the Effectiveness of Visual Encodings},
	author       = {Kim, Younghoon and Heer, Jeffrey},
	year         = 2018,
	journal      = {Comput. Graph. Forum},
	volume       = 37,
	number       = 3,
	pages        = {157--167},
	doi          = {10.1111/cgf.13409}
}

@inproceedings{wu21:lq2,
	title        = {Learning to Automate Chart Layout Configurations Using Crowdsourced Paired Comparison},
	author       = {Wu, Aoyu and Xie, Liwenhan and Lee, Bongshin and Wang, Yun and Cui, Weiwei and Qu, Huamin},
	year         = 2021,
	booktitle    = {ACM Human Factors in Computing Systems},
	location     = {Yokohama, Japan},
	publisher    = {ACM},
	address      = {New York, NY, USA},
	series       = {CHI '21},
    pages        = {14:1--14:13},
	doi          = {10.1145/3411764.3445179}
}

@inproceedings{hu19:vizml,
	title        = {VizML: A Machine Learning Approach to Visualization Recommendation},
	author       = {Hu, Kevin and Bakker, Michiel A. and Li, Stephen and Kraska, Tim and Hidalgo, C\'{e}sar},
	year         = 2019,
	booktitle    = {ACM Human Factors in Computing Systems},
	location     = {Glasgow, Scotland Uk},
	publisher    = {ACM},
	address      = {New York, NY, USA},
	series       = {CHI '19},
	pages        = {128:1--128:12},
	doi          = {10.1145/3290605.3300358}
}

@article{li2022:kg4vis,
	title        = {KG4Vis: A Knowledge Graph-Based Approach for Visualization Recommendation},
	author       = {Li, Haotian and Wang, Yong and Zhang, Songheng and Song, Yangqiu and Qu, Huamin},
	year         = 2022,
	journal      = {IEEE Trans. Vis. Comput. Graph.},
	volume       = 28,
	number       = 1,
	pages        = {195--205},
	doi          = {10.1109/TVCG.2021.3114863}
}

@article{pandey2023:genorec,
	title        = {GenoREC: A Recommendation System for Interactive Genomics Data Visualization},
	author       = {Pandey, Aditeya and L'Yi, Sehi and Wang, Qianwen and Borkin, Michelle A. and Gehlenborg, Nils},
	year         = 2023,
	journal      = {IEEE Trans. Vis. Comput. Graph.},
	volume       = 29,
	number       = 1,
	pages        = {570--580},
	doi          = {10.1109/TVCG.2022.3209407}
}

@article{zhang2024:adavis,
	title        = {AdaVis: Adaptive and Explainable Visualization Recommendation for Tabular Data},
	author       = {Zhang, Songheng and Li, Haotian and Qu, Huamin and Wang, Yong},
	year         = 2024,
	journal      = {IEEE Trans. Vis. Comput. Graph.},
	volume       = 30,
	number       = 9,
	pages        = {5923--5938},
	doi          = {10.1109/TVCG.2023.3316469}
}

@inproceedings{li2022:structure,
	title        = {Structure-aware Visualization Retrieval},
	author       = {Li, Haotian and Wang, Yong and Wu, Aoyu and Wei, Huan and Qu, Huamin},
	year         = 2022,
	booktitle    = {ACM Human Factors in Computing Systems},
	location     = {New Orleans, LA, USA},
	publisher    = {ACM},
	series       = {CHI '22},
    pages        = {409:1--409:14},
	doi          = {10.1145/3491102.3502048}
}

@article{lyi2022:gosling,
	title        = {Gosling: A Grammar-based Toolkit for Scalable and Interactive Genomics Data Visualization},
	author       = {LYi, Sehi and Wang, Qianwen and Lekschas, Fritz and Gehlenborg, Nils},
	year         = 2022,
	journal      = {IEEE Transactions on Visualization and Computer Graphics},
	volume       = 28,
	number       = 1,
	pages        = {140--150},
	doi          = {10.1109/TVCG.2021.3114876}
}

@article{nusrat2019:genomic,
	title        = {Tasks, Techniques, and Tools for Genomic Data Visualization},
	author       = {Nusrat, S. and Harbig, T. and Gehlenborg, N.},
	year         = 2019,
	journal      = {Computer Graphics Forum},
	volume       = 38,
	number       = 3,
	pages        = {781--805},
	doi          = {10.1111/cgf.13727}
}

@article{kim2025:augmentation,
	title        = {Data Augmentation for Visualization Design Knowledge Bases},
	author       = {Kim, Hyeok and Heer, Jeffrey},
	year         = 2025,
	journal      = {IEEE Transactions on Visualization and Computer Graphics},
	note         = {\url{https://www.arxiv.org/abs/2508.02216}. To appear},
}

@article{theng2024:feature,
	title        = {Feature selection techniques for machine learning: a survey of more than two decades of research},
	author       = {Theng, Dipti and Bhoyar, Kishor K.},
	year         = 2024,
	month        = {Mar},
	day          = {01},
	journal      = {Knowledge and Information Systems},
	volume       = 66,
	number       = 3,
	pages        = {1575--1637},
	doi          = {10.1007/s10115-023-02010-5}
}

@article{maryam2018:filter,
	title        = {OSFSMI: Online stream feature selection method based on mutual information},
	author       = {Maryam Rahmaninia and Parham Moradi},
	year         = 2018,
	journal      = {Applied Soft Computing},
	volume       = 68,
	pages        = {733--746},
	doi          = {10.1016/j.asoc.2017.08.034}
}

@article{sanz2018:wrapper,
	title        = {SVM-RFE: selection and visualization of the most relevant features through non-linear kernels},
	author       = {Sanz, Hector and Valim, Clarissa and Vegas, Esteban and Oller, Josep M. and Reverter, Ferran},
	year         = 2018,
	journal      = {BMC Bioinformatics},
	volume       = 19,
	number       = 1,
	pages        = 432,
	doi          = {10.1186/s12859-018-2451-4}
}

@article{pes2020:embedded,
	title        = {Ensemble feature selection for high-dimensional data: a stability analysis across multiple domains},
	author       = {Pes, Barbara},
	year         = 2020,
	journal      = {Neural Computing and Applications},
	volume       = 32,
	number       = 10,
	pages        = {5951--5973},
	doi          = {10.1007/s00521-019-04082-3}
}

@inproceedings{khalid2014:feature,
	title        = {A survey of feature selection and feature extraction techniques in machine learning},
	author       = {Khalid, Samina and Khalil, Tehmina and Nasreen, Shamila},
	year         = 2014,
	booktitle    = {2014 Science and Information Conference},
	pages        = {372--378},
	doi          = {10.1109/SAI.2014.6918213}
}

@article{wickham:ggplot22010,
	title        = {A Layered Grammar of Graphics},
	author       = {Hadley Wickham},
	year         = 2010,
	journal      = {Journal of Computational and Graphical Statistics},
	publisher    = {American Statistical Association, Taylor \& Francis, Ltd., Institute of Mathematical Statistics, Interface Foundation of America},
	volume       = 19,
	number       = 1,
	pages        = {3--28},
	doi          = {10.1198/jcgs.2009.07098},
	url          = {http://www.jstor.org/stable/25651297}
}

@article{satyanarayan:vega-lite2017,
	title        = {Vega-Lite: A Grammar of Interactive Graphics},
	author       = {Arvind Satyanarayan AND Dominik Moritz AND Kanit Wongsuphasawat AND Jeffrey Heer},
	year         = 2017,
	journal      = {IEEE Trans. Visualization \& Comp. Graphics (Proc. InfoVis)},
	url          = {https://doi.org/10.1109/TVCG.2016.2599030}
}

@inproceedings{satyanarayan:vega2016,
	title        = {Reactive Vega: A Streaming Dataflow Architecture for Declarative Interactive Visualization},
	author       = {Arvind Satyanarayan AND Ryan Russell AND Jane Hoffswell AND Jeffrey Heer},
	year         = 2016,
	booktitle    = {IEEE Trans. Visualization \& Comp. Graphics (Proc. InfoVis '15)},
	url          = {https://doi.org/10.1109/TVCG.2015.2467091}
}

@inproceedings{albers2014:task,
	title        = {Task-driven evaluation of aggregation in time series visualization},
	author       = {Albers, Danielle and Correll, Michael and Gleicher, Michael},
	year         = 2014,
	booktitle    = {Proceedings of the SIGCHI Conference on Human Factors in Computing Systems},
	location     = {Toronto, Ontario, Canada},
	publisher    = {Association for Computing Machinery},
	address      = {New York, NY, USA},
	series       = {CHI '14},
	pages        = {551–560},
	doi          = {10.1145/2556288.2557200}
}

@misc{genocat,
    title        = {GenoCAT},
    note         = {\url{http://genocat.tools/}. Last accessed Sep 6, 2025},
    author       = {},
    year         = {n.d.}
}

@misc{agv,
    title        = {Awesome Genome Visualization},
    note         = {\url{https://cmdcolin.github.io/awesome-genome-visualization/?latest=true}. Last accessed Sep 6, 2025},
    author       = {},
    year         = {n.d.}
}

@article{qu2018:multiview,
	title        = {Keeping Multiple Views Consistent: Constraints, Validations, and Exceptions in Visualization Authoring},
	author       = {Qu, Zening and Hullman, Jessica},
	year         = 2018,
	journal      = {IEEE Transactions on Visualization and Computer Graphics},
	volume       = 24,
	number       = 1,
	pages        = {468--477},
	doi          = {10.1109/TVCG.2017.2744198}
}

@article{walker2002:magnitude,
	title        = {Magnitude estimation of conceptual data dimensions for use in sonification.},
	author       = {Walker, Bruce N},
	year         = 2002,
	journal      = {Journal of experimental psychology: Applied},
	publisher    = {American Psychological Association},
	volume       = 8,
	number       = 4,
	pages        = 211
}

@article{walker2007:consistency,
	title        = {Consistency of magnitude estimations with conceptual data dimensions used for sonification},
	author       = {Walker, Bruce N},
	year         = 2007,
	journal      = {Applied Cognitive Psychology: The Official Journal of the Society for Applied Research in Memory and Cognition},
	publisher    = {Wiley Online Library},
	volume       = 21,
	number       = 5,
	pages        = {579--599}
}

@article{mcnutt2025:colorlinter,
	title        = {Mixing Linters with GUIs: A Color Palette Design Probe},
	author       = {McNutt, Andrew and Stone, Maureen C. and Heer, Jeffrey},
	year         = 2025,
	journal      = {IEEE Transactions on Visualization and Computer Graphics},
	volume       = 31,
	number       = 1,
	pages        = {327--337},
	doi          = {10.1109/TVCG.2024.3456317}
}

@article{chen2022:vizlinter,
	title        = {VizLinter: A Linter and Fixer Framework for Data Visualization},
	author       = {Chen, Qing and Sun, Fuling and Xu, Xinyue and Chen, Zui and Wang, Jiazhe and Cao, Nan},
	year         = 2022,
	journal      = {IEEE Transactions on Visualization and Computer Graphics},
	volume       = 28,
	number       = 1,
	pages        = {206--216},
	doi          = {10.1109/TVCG.2021.3114804}
}

@article{srinivasan2025:dashboard,
	title        = {From Dashboard Zoo to Census: A Case Study With Tableau Public},
	author       = {Srinivasan, Arjun and Purich, Joanna and Correll, Michael and Battle, Leilani and Setlur, Vidya and Crisan, Anamaria},
	year         = 2025,
	journal      = {IEEE Transactions on Visualization and Computer Graphics},
	volume       = 31,
	number       = 9,
	pages        = {6085--6099},
	doi          = {10.1109/TVCG.2024.3490259}
}

@article{heer2010:zoo,
	title        = {A tour through the visualization zoo: A survey of powerful visualization techniques, from the obvious to the obscure},
	author       = {Heer, Jeffrey and Bostock, Michael and Ogievetsky, Vadim},
	year         = 2010,
	journal      = {Queue},
	publisher    = {ACM},
	volume       = 8,
	number       = 5,
	pages        = {20--30},
	doi          = {10.1145/1794514.1805128}
}

@inproceedings{hofman2020:cipi,
	title        = {How Visualizing Inferential Uncertainty Can Mislead Readers About Treatment Effects in Scientific Results},
	author       = {Hofman, Jake M. and Goldstein, Daniel G. and Hullman, Jessica},
	year         = 2020,
	booktitle    = {Proceedings of the 2020 CHI Conference on Human Factors in Computing Systems},
	publisher    = {ACM},
	series       = {CHI '20},
	doi          = {10.1145/3313831.3376454}
}

@article{battle2024:pyxis,
	title        = {What Do We Mean When We Say “Insight”? A Formal Synthesis of Existing Theory},
	author       = {Battle, Leilani and Ottley, Alvitta},
	year         = 2024,
	journal      = {IEEE Transactions on Visualization and Computer Graphics},
	volume       = 30,
	number       = 9,
	pages        = {6075--6088},
	doi          = {10.1109/TVCG.2023.3326698}
}

@inproceedings{kim:erie2023,
	title        = {Erie: a Declarative Grammar for Data Sonification},
	author       = {Kim, Hyeok and Kim, Yea-Seul and Hullman, Jessica},
	year         = 2024,
	booktitle    = {ACM Proc. CHI},
	doi          = {10.1145/3613904.3642442}
}

@article{narechania2021:nl4dv,
	title        = {NL4DV: A Toolkit for Generating Analytic Specifications for Data Visualization from Natural Language Queries},
	author       = {Narechania, Arpit and Srinivasan, Arjun and Stasko, John},
	year         = 2021,
	journal      = {IEEE Transactions on Visualization and Computer Graphics},
	volume       = 27,
	number       = 2,
	pages        = {369--379},
	doi          = {10.1109/TVCG.2020.3030378}
}

@inproceedings{lin2020:dziban,
	title        = {Dziban: Balancing Agency \& Automation in Visualization Design via Anchored Recommendations},
	author       = {Lin, Halden and Moritz, Dominik and Heer, Jeffrey},
	year         = 2020,
	booktitle    = {Proceedings of the 2020 CHI Conference on Human Factors in Computing Systems},
	publisher    = {ACM},
	series       = {CHI '20},
	pages        = {1–12},
	doi          = {10.1145/3313831.3376880}
}

@article{kale2024:evm,
	title        = {EVM: Incorporating Model Checking into Exploratory Visual Analysis},
	author       = {Kale, Alex and Guo, Ziyang and Qiao, Xiao Li and Heer, Jeffrey and Hullman, Jessica},
	year         = 2024,
	journal      = {IEEE Transactions on Visualization and Computer Graphics},
	volume       = 30,
	number       = 1,
	pages        = {208--218},
	doi          = {10.1109/TVCG.2023.3326516}
}

\appendix

\end{document}